\documentclass{aa}  
\usepackage{graphicx}
\usepackage{subcaption}
\usepackage{comment}
\usepackage{amsmath}
\usepackage{xspace}
\usepackage{braket}
\usepackage{xcolor}
\usepackage{hyperref}

\renewcommand*{\corrauth}[1]{\thanks{{#1}}}
\makeatletter
\renewcommand*\aa@titlefont{\LARGE\bfseries\boldmath}
\makeatother
\newcommand{\hlinesp}{\hline\rule{0pt}{15pt}}
\newcommand{\msun}{{\rm M}_\odot}
\newcommand{\sevn}{\textsc{sevn}\xspace}

\definecolor{tiffany}{RGB}{79, 166, 158}

\newcommand{\parsec}{\textsc{parsec}\xspace}
\newcommand{\parseci}{\textsc{parsec-i}\xspace}
\newcommand{\parsecii}{\textsc{parsec-ii}\xspace}

\begin{document}

\title{The impact of stellar binaries and star cluster dynamics on pair-instability supernovae}

\author{Francesco Gabrielli\inst{1,2}\corrauth{francesco.gabrielli@physics.uu.se}
    \and Cristiano Ugolini\inst{3,4,5}\email{cristiano.ugolini@gssi.it}
    \and Lavinia Paiella\inst{3,4,5}\email{lavinia.paiella@gssi.it}
    \and Benedetta Mestichelli\inst{3,4,5,6}\email{benedetta.mestichelli@gssi.it}
    \and Manuel Arca Sedda\inst{3,4,5}\email{manuel.arcasedda@gssi.it}
    \and Lumen Boco\inst{6}\email{lumen.boco@uni-heidelberg.de}
    \and Kendall Shepherd\inst{7}\email{kendall.shepherd@unipd.it}
    \and Giuliano Iorio\inst{8}\email{giuliano.iorio@icc.ub.edu}
    \and Guglielmo Costa\inst{7,9}\email{guglielmo.costa.astro@gmail.com}
    \and Giovanni Gandolfi\inst{5}\email{giovanni.gandolfi@inaf.it}
    \and Andrea Lapi\inst{10}\email{lapi@sissa.it}
    \and Erik Zackrisson\inst{1}\email{erik.zackrisson@physics.uu.se}
    \and Thomas Nordlander\inst{1}\email{thomas.nordlander@physics.uu.se}
    \and Alessandro Bressan\inst{10}\email{sbressan@sissa.it}
    \and Mario Spera\inst{2,5,10}\corrauth{mspera@sissa.it}
    }

\institute{Department of Physics and Astronomy, Uppsala University, Box 516, SE-751 20 Uppsala, Sweden
\and National Institute for Nuclear Physics - INFN, Sezione di Trieste, I-34127 Trieste, Italy
\and Gran Sasso Science Institute, Via F. Crispi 7, L’Aquila, I-67100, Italy
\and INFN - Laboratori Nazionali del Gran Sasso, I-67100 Assergi, Italy
\and INAF - Osservatorio Astronomico di Roma, I-00040 Monte Porzio Catone (Rome), Italy
\and Institut für Theoretische Astrophysik, Zentrum für Astronomie, Universität Heidelberg, Albert Ueberle Str. 2, D-69120 Heidelberg, Germany
\and Dipartimento di Fisica e Astronomia “Galileo Galilei,” Università di Padova, Vicolo dell’Osservatorio 3, Padova, Italy
\and Departament de Física Quàntica i Astrofísica, Institut de Ciències del Cosmos, Universitat de Barcelona, Martí i Franquès 1,E-08028 Barcelona, Spain
\and INAF-Osservatorio Astronomico di Padova, Vicolo dell’Osservatorio 5, Padova, Italy
\and International School for Advanced Studies (SISSA), Via Bonomea 265, I-34136 Trieste, Italy}

\date{Received XXX}

  \abstract
   {}
   {Pair-instability supernovae (PISNe) are among the most luminous transients in the Universe. However, they have never been confidently observed. Solving this puzzle would have key implications for several astrophysical topics, including galaxy chemical enrichment, the interpretation of gravitational waves from binary black hole mergers, and the nature of red dropout sources seen by JWST. With this aim, we present the first in-depth study of PISN occurrence in binary stars, both in isolation and in dense star clusters.} 
   {We employ the \sevn code, with state-of-the-art \textsc{parsec} stellar tracks, to evolve a suite of 35 synthetic binary populations, including variations on formation channels, cluster properties, and upper limit of the stellar initial mass function. Combining with an up-to-date, semi-empirical determination of the metallicity-dependent star formation rate density, we provide new results on the cosmic rate of PISNe.}
   {We find that binary interactions can boost the PISN rate by up to threefold, relative to single stars, whereas binary hardening can either enhance or suppress PISN production, depending on whether the progenitors are primordial or dynamically formed. Moreover, we showcase how our comprehensive framework for the cosmic PISN rate can be used to constrain uncertain aspects of stellar and galaxy evolution models, via comparison with observations, including the recipes for stellar-wind mass loss in very-massive stars, and the galaxy metallicity distribution throughout the Universe.}
   {}


\maketitle

\section{Introduction}\label{sec:intro}

Pair-instability supernovae (PISNe) are expected to mark the final fate of low-metallicity, very massive stars (VMSs, \citealp{Fowler_1964,Barkat_1967,Bisnovatyi_Kogan_1967,Rakavy_1967,Fraley_1968,Heger_2002,Heger_2003}). During oxygen (O) burning in the core, or at the end of core carbon (C) burning, photons can become energetic enough to produce electron-positron pairs, due to temperatures up-to $10^9$ K and densities $\gtrsim 100\ \rm g\ cm^{-3}$. This removes central radiation pressure, leading the star into a phase of runaway collapse. Ultimately, explosive O and silicon (Si) burning releases energies of order $10^{52}$--$10^{53}$ erg, sufficient to completely disrupt the star and leave no compact remnant \citep[e.g.,][]{Heger_2002,Heger_2003}. This complete disruption is one of the main ingredients behind the upper mass gap in the black hole (BH) mass spectrum, i.e. the predicted dearth of BHs with masses between $\sim50$ and 120 $\msun$ \citep[e.g.,][]{Spera_2017,Farmer_2019}. Moreover, it sets them apart from pulsational-PISNe, that involve less-massive progenitors, and eventually lead to the formation of a BH. The radioactive decay of the nickel isotope $^{56}$Ni, synthesized in large amounts up to $50$--$60\,\msun$ during the PISN explosion, can produce luminosities up to $10^2$ times higher than those of typical core-collapse supernovae (CCSNe, \citealp{Scannapieco_2003,Kasen_2011,Dessart_2012,Whalen_2013,Kozyreva_2014,Kozyreva_2014_1,Kozyreva_2016,Jerkstrand_2015,Smidt_2015,Gilmer_2017,Hartwig_2018,Chatzopoulos_2019}). Despite such high luminosities, PISNe have never been confidently observed, with only candidate identifications so far \citep{Woosley_2007,Gal_Yam_2009,Quimby_2011,Cooke_2012,Gal_Yam_2012,Kozyreva_2015,Lunnan_2016,Kozyreva_2018,Gomez_2019,Mazzali_2019,Nicholl_2020,Ferrara_2026,Jeon_2026}. Recent candidates include SN 2018ibb, at $z=0.166$ \citep{Schulze_2024}, SN 2023vbw, at $z=0.088$ \citep{Hiramatsu_2026}, and the proposed PISN interpretation of the ultra-early James Webb Space Telescope (JWST) source \textit{Capotauro} at $z\sim 15$ \citep{Gandolfi_2026,Ferrara_2026,Jeon_2026}. However, none of these candidates has been confirmed yet.

Most massive stars, and likely also VMSs, are expected to be born in binary or higher-order multiple systems, with binary fractions approaching unity for stars $\gtrsim20\,\msun$ \citep{Sana_2012,Moe_2017,Winters_2019,Offner_2022}. Therefore, a large fraction of PISN progenitors should experience binary evolution. Binary interactions can critically affect stellar evolution, and especially the onset of pair instability \citep[e.g.,][]{Vigna_Gomez_2019, Marchant_2019,Di_Carlo_2019,Di_Carlo_2020,Renzo_2020a,Zapartas_2021}. Furthermore, it is still not clear whether the VMSs observed in the local Universe were born with such high masses, or if they originated from stellar mergers, leaving the question regarding the upper limit of the stellar initial mass function (IMF) open (\citealp[e.g.,][]{Crowther_2010,Evans_2010,Vink_2015,Schneider_2018,Brands_2022,Crowther_2023}).  

Binaries can reside in a wide variety of environments, from the low-density galactic regions (``field"), to dense star clusters. Observations of the local Universe show that most massive stars form in associations or clusters \citep{Lada_2003,Portegies_Zwart_2010}. In dense environments, binary evolution can be altered by dynamical encounters including binary hardening, exchanges, collisions, and disruptions \citep[e.g.,][]{Heggie_1975,Benacquista_2006,binney_tremaine_2008,Marin_Pina_2026,Rastello_2026}. Star clusters host both primordial binaries, whose components are born together, and dynamically assembled binaries, which form through encounters between initially unrelated stars. These processes may either promote stellar mergers and mass transfer, or remove potential PISN progenitors from the relevant mass range. 

A realistic estimate of the PISN rate therefore requires accounting for both binary evolution and the environments in which massive stars form. This is important for several open problems, including the origin of the upper mass gap in the BH mass spectrum \citep[e.g.,][]{Spera_2017,Farmer_2019,LVK_2025,Tong_2026}, the formation and maximum mass of VMSs \citep[e.g.,][]{Crowther_2010,Evans_2010,Vink_2015,Schneider_2018,Brands_2022,Crowther_2023}, the interpretation of red dropout sources detected with JWST \citep[e.g.,][]{Whalen_2012,Hartwig_2018,Gandolfi_2026,Ferrara_2026,Jeon_2026}, and the chemical enrichment of galaxies \citep[e.g.,][]{Ricotti_2004,Matteucci_2005,Romano_2020,Goswami_2021,Goswami_2022,Xing_2023,Koutsouridou_2024,Goswami_2024,Vanni_2024}.

However, a systematic population-synthesis study of PISN production across isolated and dynamically processed binary populations is still missing.

In this paper, we present such a study, focusing on how binary interactions and dynamical processes in dense star clusters affect PISN production relative to single-star evolution. We employ the population synthesis code \sevn \citep{Spera_2015,Iorio_2023}, with \parsec stellar tracks \citep{Bressan_2012,Costa_2025}, to evolve synthetic populations of single and binary stars, and model the secular hardening of binaries due to binary--single encounters in dense stellar environments. Moreover, by combining these results with an up-to-date, semi-empirical determination of the metallicity-dependent star formation history, we build a comprehensive theoretical framework for the cosmic PISN rate, accounting for multiple progenitors and galactic environments, and show how comparisons with observations can constrain uncertain ingredients of stellar and galaxy evolution.

The paper is structured as follows. In Sec.~\ref{sec:methods}, we present the methods followed in this work, and the variations we consider in the model assumptions. In Sec.~\ref{sec:results}, we present our results, that we discuss in Sec.~\ref{sec:discussion}. We finally draw our conclusions in Sec.~\ref{sec:conclusions}.

\section{Methods} \label{sec:methods}

Throughout this work, we assume a flat $\Lambda$CDM cosmology from \cite{Planck_2020}, with parameters $\Omega_M=0.32$, $\Omega_b=0.05$, $H_0=67\rm \:km\:s^{-1}\:Mpc^{-1}$. We adopt a standard Kroupa IMF \citep{Kroupa_2001}, from $0.1\:\msun$ to a varying upper mass limit. Following \cite{Caffau_2010}, we use $Z_{\odot}=0.0153$ for the solar metallicity, and $12+\rm \log(O/H)_{\odot}=8.76$ for the solar oxygen abundance. This allows us to define the gas-phase metallicity as $\log Z\simeq 12+\log (\rm O/H)-10.58$, based on the O abundance. See Sec.~ \ref{sec:discussion_other_interactions} for a discussion about using the iron (Fe) abundance instead.

\subsection{\sevn and binary hardening}

We simulate the evolution of single and binary stars with the population synthesis code \sevn\footnote{In this work, we use the \sevn version V 2.10.0 (commit ab9b047b). SEVN is publicly available at the gitlab repository https://gitlab.com/sevncodes/sevn.} \citep{Spera_2015,Spera_2017,Spera_2019,Iorio_2023}. \sevn interpolates pre-computed stellar evolution tracks and couples them to semi-analytic prescriptions for binary evolution. This allows us to follow stellar populations from the zero-age main sequence (ZAMS) to compact object formation or complete disruption. We adopt one of the most recent sets of tracks computed with the \parsec stellar evolution code \citep{Bressan_2012,Costa_2019,Costa_2021,Nguyen_2022,Costa_2025,Nguyen_2025}, corresponding to the version used in \cite{Iorio_2023}, and referred to as \parsecii in \cite{Gabrielli_2024a}.

These tracks are non-rotating and cover metallicities from $Z=10^{-11}$ to $3\times 10^{-2}$. The impact of rotation is not included in this work and is discussed as a caveat in Sec.~\ref{sec:discussion_other_interactions}.

We classify stars as PISN progenitors following the prescriptions of \cite{Spera_2017} and \cite{Mapelli_2020}. In particular, stars with final helium-core masses in the range $64\,\msun\leq M_{\rm He,fin}\leq 135\,\msun$ undergo complete disruption through a PISN, leaving no compact remnant. We keep the main binary-evolution prescriptions fixed throughout the paper. For common envelope (CE) evolution, we adopt the formalism of \cite{Hurley_Tout_Pols_2002}, with $\alpha_{\rm CE}=3$ and $\lambda_{\rm CE}$ computed as in the \textsc{BSE} code \citep{Claeys_2014}. For Roche-lobe overflow (RLO), we follow the implementation of \cite{Iorio_2023}, based on \cite{Hurley_Tout_Pols_2002}. In particular, we adopt a mass-transfer efficiency $f_{\rm MT}=0.5$, with the non-accreted material lost from the system. The stability of mass transfer is assessed via $q_{\rm crit}$, i.e. a critical value for the donor-to-accretor mass ratio, as described in \cite{Iorio_2023}. Moreover, SEVN identifies stellar collisions when the two binary components come in contact at periastron. Stellar mergers can instead arise from a CE phase, RLO, a stellar collision, a combination, or none of the above. In case of a merger, the total, He- and CO-core masses of the two components are summed, and the remnant star is assigned the stellar phase and life percentage of the more-evolved component.

To model the effect of dense stellar environments, we use the binary-hardening option implemented in \sevn. This prescription describes the secular evolution of binary semi-major axes and eccentricities due to repeated binary--single encounters in star clusters. It does not follow individual few-body encounters, exchanges, ejections, or binary--binary interactions, but it captures the average hardening of hard binaries. We refer the reader to Sec.~ \ref{sec:discussion_other_interactions} for an explanation of why this approach is best suited for this work. The rates of decrease of the semi-major axis and increase of the eccentricity are given by \citep{Heggie_1975}:
\begin{equation}
\frac{da}{dt}=-2\pi\xi\frac{G\rho}{\sigma}a^2,
\end{equation}
\begin{equation}
\frac{de}{dt}=2\pi\xi\kappa\frac{G\rho}{\sigma}.
\end{equation}
$\xi$ and $\kappa$ are dimensionless parameters calibrated on direct N-body and scattering experiments \citep{Quinlan_1996,Miller_Hamilton_2002,Sesana_2006}. Following \citealp{Mapelli_2021a,Mapelli_2022,Vaccaro_2024, Torniamenti_2024}, we fix $\xi=3$, $\kappa = 0.1$, and assume an average mass of stellar perturbers of $m_{\star}=1\:\msun$.

\subsection{Star cluster properties}\label{sec:methods_clusters}

\begin{table}
\centering
\captionof{table}{Properties of the star clusters considered in this work. From left to right, we report the cluster label, total stellar mass, half-mass radius, central mass density and velocity dispersion inferred from a Plummer model, and the corresponding binary-hardening rate defined in Eq.~\ref{eq:rhard}.}
\resizebox{\columnwidth}{!}{%
\begin{tabular} { c c c c c c }
\hline\vspace{-2mm}\\
cluster & $M_{\rm cl}$ & $R_{\rm h,cl}$ & $\rho_{\rm c}$ & $\sigma_{\rm c}$ & $\mathcal{R}_{\rm hard}$\\
label & $[\msun]$ & $[\rm pc]$ & $[\msun\:\rm pc^{-3}]$ & $[\rm km\:s^{-1}]$ & $[\rm pc^{-1}\:Myr^{-1}]$\\\vspace{-2mm}\\
\hline\vspace{-2mm}\\
G1 & $10^5$ & $1$ & $5.2\times 10^4$ & $9.7$ & $4.5\times 10^2$ \\
G2 & $10^6$ & $1$ & $5.2\times 10^5$ & $30.5$ & $1.4\times 10^3$ \\
G3 & $10^5$ & $5$ & $4.2\times 10^2$ & $4.3$ & $8.1\times 10^0$ \\
G4 & $10^6$ & $5$ & $4.2\times 10^3$ & $13.7$ & $2.5\times 10^1$ \\
Y1 & $10^4$ & $0.2$ & $6.6\times 10^5$ & $6.8$ & $8.0\times 10^3$ \\
Y2 & $10^5$ & $0.2$ & $6.6\times 10^6$ & $21.6$ & $2.5\times 10^4$ \\
Y3 & $10^4$ & $1$ & $1.12\times 10^3$ & $1.7$ & $1.4\times 10^2$ \\
N1 & $10^7$ & $1$ & $5.2\times 10^6$ & $96.5$ & $4.5\times 10^3$ \\
N2 & $10^7$ & $5$ & $4.2\times 10^4$ & $43.2$ & $8.1\times 10^1$
%\hline
\label{tab:clusters}
\end{tabular}
}
\end{table}

To sample different regions of the cluster mass--radius plane, and quantify how the efficiency of binary hardening changes across them, we consider three broad classes of dense stellar environments, for a total of nine cluster models: globular-cluster-like (GC) models (G1--G4), with properties compatible with those in the Milky Way and local galaxies; young-star-cluster-like (YSC) models (Y1--Y3); and nuclear-star-cluster-like (NSC) models (N1, N2)\footnote{Compact, massive star clusters with $\sim 10^5$--$10^6\ M_\odot$ and sizes $R_\mathrm{hl, cl}\lesssim 1$ pc have recently been identified through JWST at $z\approx 6$--10 \citep[for a compilation, see][]{Messa26}. These systems occupy a region of the mass--radius plane comparable to our most compact YSC-like and GC-like models (Y2 and G2). We therefore expect our calculations to provide a useful reference for this class of environments, although a dedicated treatment of their formation and evolution is beyond the scope of this work.}. For each cluster model, we assign a total stellar mass, $M_{\rm cl}$, and a half-mass radius, $R_{\rm h,cl}$, and compute the  corresponding central mass density, $\rho_{\rm c}$, and central (one-dimensional) velocity dispersion, $\sigma_{\rm c}$, assuming a Plummer density profile \citep{Plummer_1911}. For a Plummer model, the scale radius is $r_{\rm p}=R_{\rm h,cl}/1.305$, and
\begin{equation}
\rho_{\rm c} = \frac{3M_{\rm cl}}{4\pi r_{\rm p}^3},\ \ \ \sigma_{\rm c}^2 = \frac{G M_{\rm cl}}{6 r_{\rm p}} .
\end{equation}
We then run \sevn with the hardening option set to these values, and summarize the resulting cluster properties in Table~\ref{tab:clusters}.

Following \cite{Heggie_1975}, the hardening rate is defined as 
\begin{equation}
\mathcal{R}_{\rm hard}= \frac{d}{dt}\left(\frac{1}{a}\right)
=2\pi\xi\frac{G\rho_c}{\sigma_c}, \left[\rm pc^{-1}\ Myr^{-1}\right].
\label{eq:rhard}
\end{equation}
This quantity sets the instantaneous hardening rate applied in our simulations, and also provides a convenient scalar for comparing the relative strength of hardening across cluster models.

We estimate the core-collapse time, $t_{\rm cc}$, from the half-mass relaxation time \citep{spitzer_1969, binney_tremaine_2008}
\begin{equation}
    t_{\rm rel} = 4.2\,\mathrm{Gyr}\,\left( \frac{15}{\ln \Lambda_{\rm c}}\right)\,\left( \frac{R_{\rm h , cl}}{4 \rm\,pc}\right)\,\sqrt{\frac{M_{\rm cl}}{10^7 \, \msun}}\,,
    \label{eq:trel}
\end{equation}
where $\ln\Lambda_{\rm c}\approx\ln(0.1\,N)$ is the Coulomb logarithm and $N=M_{\rm cl}/m_\star$ is the number of stars, and adopt the approximate relation \citep{portegies_zwart_runaway_2002}
\begin{equation}
    t_{\rm cc} = 0.2 \ t_{\rm rel} \ .
\end{equation}

\begin{figure*}[ht!]
\centering
\includegraphics[scale=0.7]{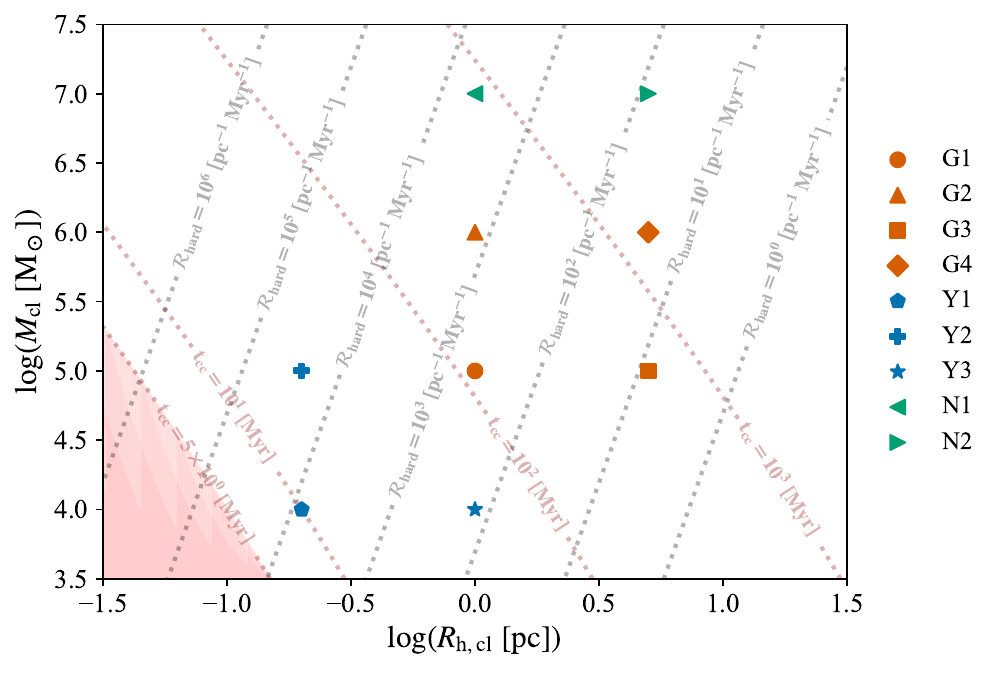}
\caption{Initial stellar masses and half-mass radii for all star clusters considered in this work. Red markers show our GCs, blue markers YSCs, and green markers NSCs. Dotted lines indicate the regions in the plane with constant binary hardening rate (grey), and cluster core-collapse timescale (red). The red shaded area indicates the region where $t_{\rm cc}\leq 5\:\rm Myr$, i.e. comparable to $t_{\rm ev}$ for most massive stellar binaries. All clusters have higher $t_{\rm cc}$ than this value.}
\label{fig:clusters_mass_radius}
\end{figure*}

Figure~\ref{fig:clusters_mass_radius} shows our cluster models in the $M_{\rm cl}$--$R_{\rm h,cl}$ plane, color-coded by class (red for GC-like, blue for YSC-like, and green for NSC-like models). Overlaid are lines of constant hardening rate $\mathcal{R}_{\rm hard}$ (grey) and constant core-collapse time $t_{\rm cc}$ (red). The shaded region marks $t_{\rm cc}\leq 5\,\rm Myr$, comparable to the evolutionary timescale $t_{\rm ev}$ of the most massive binaries that produce PISNe. All our cluster models lie above this threshold, i.e. they have $t_{\rm cc}\gtrsim t_{\rm ev}$.

In our calculations, we use $t_{\rm cc}$ only to limit the duration over which the initial hardening rate is applied: after $t_{\rm cc}$, we switch off the hardening prescription. This simplified choice avoids extrapolating the initial central density and velocity dispersion beyond the early cluster-evolution phase. As noted above, it has little impact on our results, since the evolution times of the massive binaries producing PISNe are typically shorter than $t_{\rm cc}$ for the cluster models considered here.

We adopt the central density and velocity dispersion as reference values for the binary--single interaction rate, since massive stars and massive binaries are expected to preferentially populate the inner regions of dense stellar systems because of mass segregation (\citealp{Chandrasekhar_1943,Spitzer_1988}, see also Sec. \ref{sec:discussion_other_interactions}).

\subsection{\sevn initial conditions}\label{sec:sevn_ic}

We generate two sets of initial conditions for the binary populations evolved with \sevn. These sets differ in the initial distributions for component masses, semi-major axis, and eccentricity, as they are meant to represent isolated binaries, and dynamically assembled hard binaries in dense stellar environments. 

The first set describes binaries that formed in isolation, and corresponds to the standard initial conditions usually adopted in \sevn (see \citealp{Iorio_2023}). We adopt it for both our isolated-binaries populations, and our models of primordial binaries in dense star clusters. This allows us to study the effect of binary hardening on PISN production, compared to pure stellar and binary evolution, starting from the same initial population. We draw the primary mass from a Kroupa IMF, $p(M_1)\propto M_1^{-2.3}$, with $M_1\geq 8\,\msun$. The secondary mass is assigned through the mass ratio $q=M_2/M_1$, drawn from $p(q)\propto q^{-0.1}$, with $q\in[q_{\rm min},1]$ and $q_{\rm min}=\max(5\,\msun/M_1,0.1)$, so that $M_2\geq 5\,\msun$. Orbital periods follow $p(\mathcal{P})\propto \mathcal{P}^{-0.55}$, where $\mathcal{P}=\log(P/\rm day) \in [0.15, 5.5]$, and eccentricities follow $p(e)\propto e^{-0.42}$, with $e \in [0, 0.9]$ \citep{Sana_2012}. The adopted cut $M_1\geq 8\,\msun$ reduces the computational cost while retaining binaries that can enter the PISN progenitor range through stellar mergers or mass transfer. We do not impose a higher cut because binary interactions can produce PISNe from systems with substantially lower initial primary masses. The correction for the missing low-mass stellar population is described in Sec.~\ref{sec:methods_pisn_efficiency}. We vary the IMF upper limit between $M_{\rm up}=150\,\msun$ and $300\,\msun$, following \cite{Gabrielli_2024a}. For the least massive YSC models (Y1 and Y3), we also consider a conservative case with $M_{\rm up}=85\,\msun$. This value is motivated by the maximum-stellar-mass -- cluster-mass relation used in \cite{Rantala_2024}, based on \cite{Weidner_Kroupa_2006} and \cite{Yan_2023}, and usually associated with optimal IMF sampling \citep{Kroupa_2013,Schulz_2015}. We use this case only as an exploratory scenario for low-mass clusters, where single stars may not reach the PISN mass range and PISNe can only be produced through binary interactions.

The second set of initial conditions is chosen to mimic hard binaries that have already been assembled and processed by dynamical interactions in star clusters. The primary masses are drawn from the same IMF as in the first set. The secondary masses are instead sampled from $p(M_2)\propto (M_1+M_2)^4$ \citep{O_Leary_2016, Torniamenti_2024}. We generate semi-major axes from a half-Gaussian distribution defined for $a<a_{\rm mean}$, with $a_{\rm mean}=0.1\,a_{\rm h}$ and $\sigma_{\rm h}=0.3\,a_{\rm mean}$. $a_{\rm h}=G(M_1+M_2)/\sigma_c^2$ is the hard-binary separation, defined as the limit above which a binary is considered soft. Finally, we extract eccentricities from a thermal distribution, $p(e)=2\:e$. This second set therefore differs from the first set mainly because it favors massive companions, hard separations, and high eccentricities, as expected for dynamically paired massive stars.

In all cluster models, we assume that the binaries evolved with this second set of initial conditions are hard and can undergo the hardening prescription described in Sec.~\ref{sec:methods_clusters}. This is a simplifying assumption, but it is appropriate for the massive binaries relevant to PISN production and so we expect this approximation to have negligible effects on our estimates of the PISN population and rates.

For each model variation, we evolve $10^7$ binaries for 15 metallicities, $Z=[2\times10^{-4},\:3\times 10^{-4},\:4\times 10^{-4},\:5\times 10^{-4},\:7\times 10^{-4},\:1\times 10^{-3},\:1.4\times 10^{-3},\:2\times 10^{-3},\:3\times 10^{-3},\:4\times 10^{-3},\:5\times 10^{-3},\:7\times 10^{-3},\:1\times 10^{-2},\:1.4\times 10^{-2},\:2\times 10^{-2}]$. Thus, each variation contains a total of $1.5\times 10^8$ simulated binaries. Combining the two sets of initial conditions with the different cluster models and IMF upper limits yields the full suite of 35 population variations explored in this work.

\subsection{PISN production efficiency}\label{sec:methods_pisn_efficiency}

We define the PISN production efficiency as the number of PISNe that are produced by a single stellar or binary population, per unit stellar mass available. For a given synthetic population with metallicity $Z$, we compute this quantity as:
\begin{equation}
\frac{dN_{\rm PISN}}{dM_{\star}}(Z)=\frac{N_{\rm PISN}(Z)}{M_{\rm sim}},
\end{equation}
where $N_{\rm PISN}(Z)$ is the total number of PISN events that are produced in the population, and $M_{\rm sim}$ is the total simulated stellar mass. We correct this quantity to account for the fact that we only simulate binaries with $M_1\geq 8\:\msun$, and $M_2\geq 5\:\msun$. Specifically, we also generate initial component mass distributions sampling $M_{\rm 1}$ and $M_{\rm 2}$ down to $0.1\ M_{\odot}$, and compute the correction factor as the ratio between the total mass generated in the two cases. We obtain $dN_{\rm PISN}/dM_{\star}$ at any metallicity by linearly interpolating between the available \sevn values in $\log Z$. Finally, we define the maximum metallicity of PISN progenitors, $Z_{\rm max}$, as the metallicity at which the interpolated $dN_{\rm PISN}/dM_{\star}$ vanishes.

\section{Results} \label{sec:results}

\subsection{Single stars and isolated binaries}\label{sec:pisncluster_results_star_binary}

Table~\ref{tab:mzams_intervals_cluster} reports the ZAMS mass ranges of PISN progenitors in our single-star and isolated-binary populations, for a representative set of metallicities. For binaries, we include PISNe produced by both primary and secondary stars. 

\begin{figure}[ht!]
\centering
\includegraphics[width=\columnwidth]{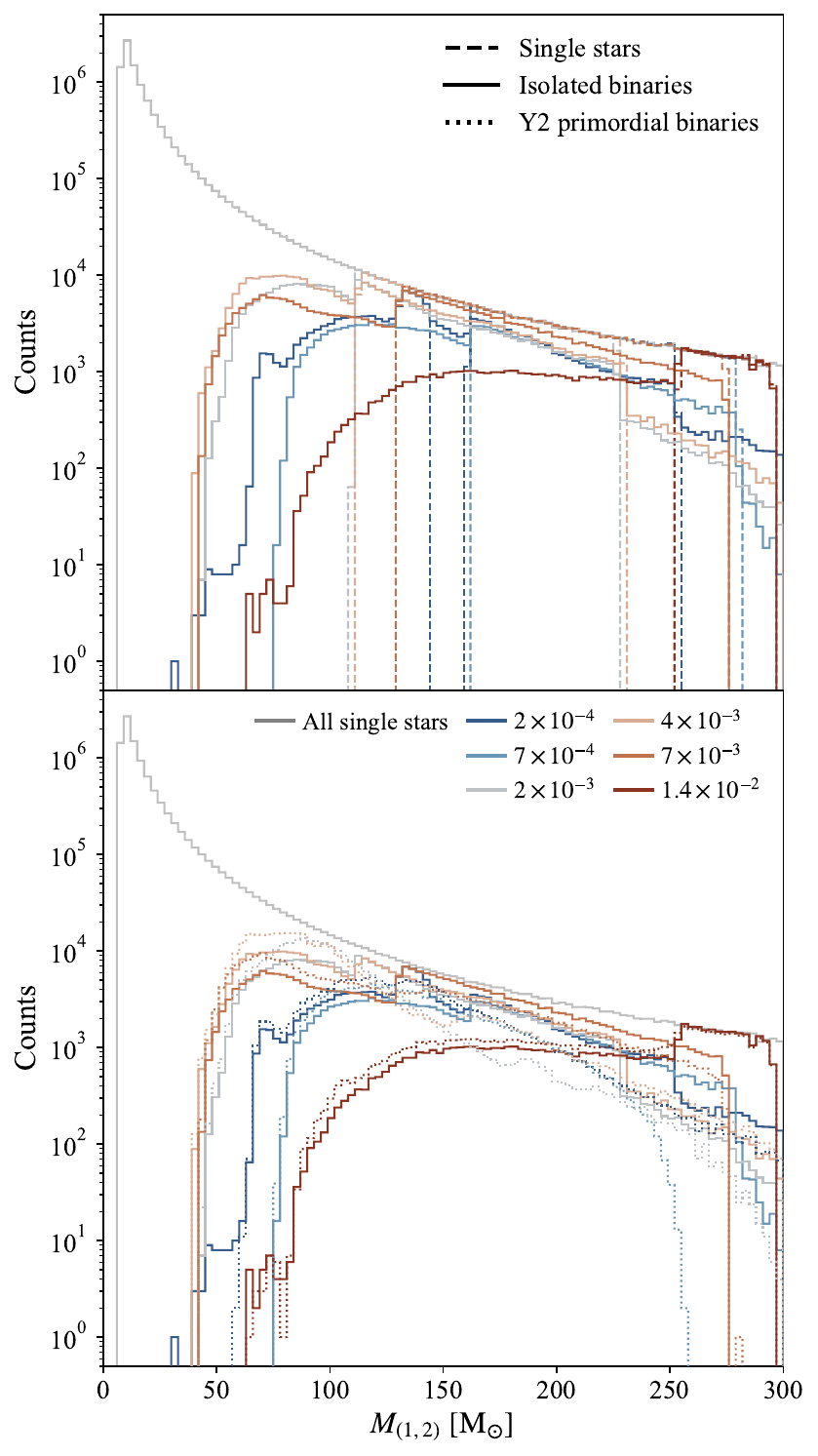}
\caption{ZAMS mass distributions of PISN progenitors for representative metallicities. The top panel compares single stars and isolated binaries. For binaries, we include PISNe produced by both primary and secondary stars. The bottom panel compares isolated binaries with primordial binaries in cluster Y2,, showing the effect of efficient hardening. The grey distribution shows the input ZAMS mass distribution of the simulated single stars and binary primaries.}
\label{fig:pisn_masses}
\end{figure}

Fig.~\ref{fig:pisn_masses} shows that binary interactions significantly broaden the initial mass range of PISN progenitors. In our single-star models, PISNe occur only for ZAMS masses above $\sim 110\,\msun$. In contrast, for binary systems PISNe are produced from components with initial masses down to $\sim 40\,\msun$. These lower-mass progenitors are produced by binary interactions, mainly stellar mergers and mass transfer, which increase the final helium-core mass of stars that would not enter the PISN regime as single stars. We discuss the relative role of mergers, mass transfer, and non-interacting binaries in Appendix \ref{sec:app_initial_properties}. Binary interactions can also move stars out of the PISN regime. A component initially inside the single-star PISN mass range may accrete enough mass to become too massive and collapse directly into an IMBH, or lose enough mass to avoid PISN. For this reason, not every binary component whose initial mass lies in the single-star PISN interval produces a PISN. In contrast, mass transfer can also reduce the mass of stars initially above the single-star PISN range, allowing them to enter the PISN regime. This explains why isolated binaries can produce PISNe from initial masses up to $\sim300\,\msun$ at most metallicities.

In the isolated-binary population, these effects produce two broad classes of PISN progenitors, visible in Fig.~\ref{fig:pisn_masses} as a high-mass component and a lower-mass tail. The first class consists of weakly interacting or non-interacting binaries. In these systems, the exploding component evolves almost as a single star, and the initial-mass distribution therefore follows the single-star PISN range. The second class consists of interacting binaries, mainly systems undergoing mergers or mass transfer. These systems populate the lower-mass tail of the distribution, below the single-star PISN range. The relative importance of the two classes depends not only on the component masses, but also on the initial semi-major axis and eccentricity. We quantify these channels in Appendix~\ref{sec:app_initial_properties}.

Fig.~\ref{fig:dn_dm_hard_singlebinary} shows the PISN production efficiency, $dN_{\rm PISN}/dM_\star$, for single stars and isolated binaries, for different IMF upper limits. For $M_{\rm up}=300\,\msun$, single stars are more efficient than isolated binaries by less than a factor of two at most metallicities. The exception is at $Z\gtrsim 10^{-2}$, where single-star PISN progenitors shift to very high initial masses, $\gtrsim250\,\msun$, and become strongly suppressed by the IMF. In this metallicity range, isolated binaries remain efficient because they can produce PISNe from lower initial masses, around $60$--$70\,\msun$. For $M_{\rm up}=150\,\msun$, isolated binaries become more efficient than single stars over most of the metallicity range. This upper-mass cut removes a large fraction of single-star PISN progenitors, especially at $2\times10^{-4}<Z<10^{-3}$ and $Z>7\times10^{-3}$, where the single-star entry mass lies above $150\,\msun$. In contrast, binaries can still produce PISNe below this limit through mergers and mass transfer. Finally, for $M_{\rm up}=85\,\msun$, single stars never reach the PISN regime in our models, while binaries can still produce PISNe through interactions.

From Fig.~\ref{fig:dn_dm_hard_singlebinary} it is also apparent that the maximum metallicity for PISN production also depends on the channel and on $M_{\rm up}$. For isolated binaries, we find PISNe up to $Z_{\rm max}=1.4\times 10^{-2}$ for all IMF upper limits considered here. Single stars reach the same $Z_{\rm max}$ only for $M_{\rm up}=300\,\msun$. For $M_{\rm up}=150\,\msun$, the single-star value decreases to $Z_{\rm max}=7\times10^{-3}$. For $M_{\rm up}=85\,\msun$, single stars produce no PISNe at any metallicity, so $Z_{\rm max}$ is not defined.

\begin{table*}
\centering
\captionof{table}{ZAMS mass ranges of PISN progenitors, for $M_{\rm up}=300\:\msun$ and representative \sevn metallicities, obtained with the \parsecii stellar evolution tracks. The first column indicates the galactic environment. The reported intervals include all stellar components that produce PISNe, including primary and secondary stars. All mass values are in solar units. Blanks indicate cases with no PISNe. The double interval at $Z=2\times 10^{-4}$ for single stars is due to the non-monotonic behaviour of the \parsecii stellar tracks at low metallicity \citep[e.g.,][]{Iorio_2023}.}
\begin{tabular}{ c c c c c c c c }
\hlinesp
\raisebox{-2ex}{env.} \hspace{-0.9cm} \rotatebox[origin=c]{35}{$\Biggm\backslash$} \hspace{-0.7cm} \raisebox{1.5ex}{\textit{Z}} & $2\times 10^{-4}$ & $7\times 10^{-4}$ & $2\times 10^{-3}$ & $4\times 10^{-3}$ & $7\times 10^{-3}$ & $1.4\times 10^{-2}$ & $2\times 10^{-2}$\\
\hline\\
\multicolumn{8}{ c }{Single stars}\\
\hlinesp
isol. & 130-143 & 162-280 & 111-227 & 112-230 & 130-275 & 254-296 & - \\ %OK
& 161-253 & & & & & & \\\\

\multicolumn{8}{ c }{Isolated binaries}\\
\hlinesp
isol. & 33-300 & 76-299 & 44-300 & 40-299 & 43-275 & 64-296 & - \\\\ %OK

\multicolumn{8}{ c }{Primordial binaries}\\
\hlinesp
G1 & 42-300 & 76-300 & 44-300 & 40-299 & 43-275 & 67-296 & 242-286 \\ %OK

G2 & 32-300 & 76-300 & 44-300 & 40-299 & 43-275 & 69-296 & 240-294 \\ %OK

G3 & 33-300 & 77-300 & 44-300 & 40-299 & 43-275 & 64-296 & - \\ %OK

G4 & 33-300 & 76-299 & 44-300 & 40-299 & 43-275 & 65-296 & - \\ %OK

Y1 & 65-85 & 76-85 & 44-85 & 40-85 & 43-85 & 61-85 & - \\ %OK

Y2 & 60-300 & 75-256 & 44-300 & 40-299 & 43-280 & 65-296 & - \\ %OK

Y3 & 65-85 & 76-85 & 44-85 & 40-85 & 43-85 & 64-85 & - \\ %OK

N1 & 33-300 & 76-300 & 44-300 & 40-299 & 43-276 & 68-296 & 229-292 \\ %OK

N2 & 33-300 & 76-300 & 44-300 & 40-299 & 43-275 & 63-296 & - \\\\ %OK

\multicolumn{8}{ c }{Dynamical binaries}\\
\hlinesp
G1 & 65-300 & 80-300 & 54-300 & 44-299 & 46-297 & 73-296 & 218-294 \\ %OK

G2 & 36-300 & 80-300 & 44-300 & 41-299 & 44-279 & 72-296 & 224-294 \\ %OK

G3 & 66-300 & 82-299 & 56-299 & 49-299 & 49-298 & 98-296 & 223-294 \\ %OK

G4 & 69-300 & 79-300 & 55-300 & 44-299 & 47-275 & 64-296 & 226-293 \\ %OK

Y1 & 65-85 & 81-85 & 49-85 & 43-85 & 46-85 & 78-84 & - \\ %OK

Y2 & 62-300 & 76-271 & 45-296 & 42-299 & 44-281 & 65-296 & - \\ %OK

Y3 & 65-85 & 81-85 & 55-85 & 46-85 & 56-85 & - & - \\ %OK

N1 & 65-300 & 75-272 & 44-298 & 40-300 & 43-276 & 62-296 & - \\ %OK

N2 & 38-300 & 77-300 & 44-300 & 41-299 & 43-276 & 64-296 & - %OK

\label{tab:mzams_intervals_cluster}
\end{tabular}
\end{table*}

\begin{figure}[ht!]
\centering
\includegraphics[width=\columnwidth]{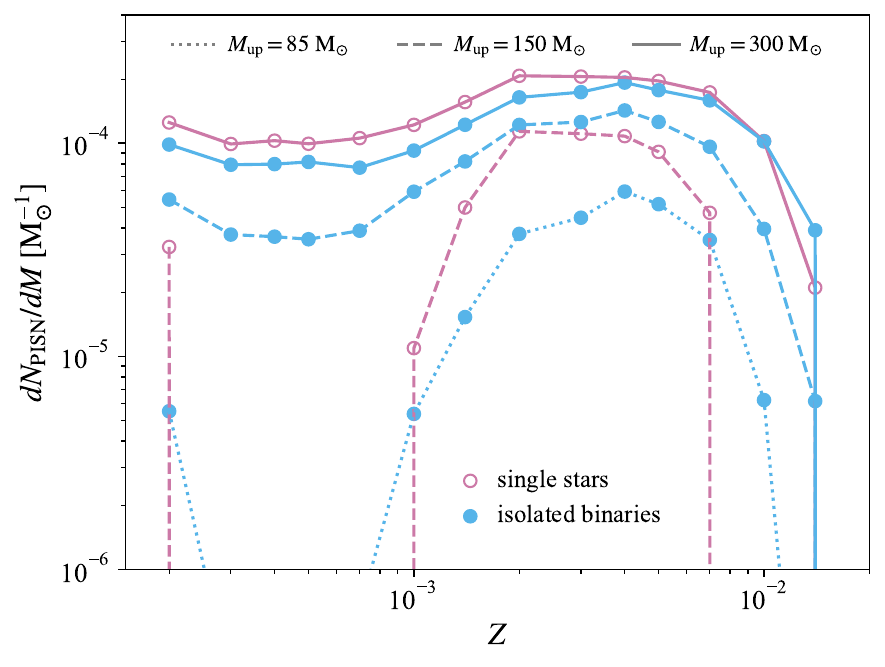}
\caption{Number of PISNe per unit star-forming mass obtained for the single stars and isolated binaries (pink and light blue lines respectively). We also show variations $M_{\rm up}\in [85,150,300]\:\msun$ (dotted, dashed, and solid lines). These results comprise PISNe arising from both primary and secondary components.}
\label{fig:dn_dm_hard_singlebinary}
\end{figure}

\subsection{Primordial binaries}\label{sec:pisncluster_results_ori_bin} 

Table~\ref{tab:mzams_intervals_cluster} reports the ZAMS mass ranges of binary PISN progenitors in dense star clusters, for both primordial and dynamically motivated initial conditions. In this section, we focus on primordial binaries subject to hardening, while the dynamically motivated models are discussed in Sec.~\ref{sec:pisncluster_results_dyn}.

For primordial binaries, the PISN progenitor mass ranges are broadly similar to those of isolated binaries. Thus, for most cluster models, hardening does not strongly change which initial stellar masses can produce PISNe. The main differences appear at the lowest and highest metallicities, and in the most compact clusters. At $Z=2\times 10^{-4}$, the lower edge of the PISN progenitor range varies between $32\,\msun$ and $65\,\msun$, compared to $33\:\msun$ for isolated binaries. The same quantity varies between $61\,\msun$ and $69\,\msun$ at $Z=1.4\times 10^{-2}$, while it is $64\:\msun$ for isolated binaries. Moreover, at $Z=7\times 10^{-4}$, cluster Y2 stands out for exhibiting an upper PISN progenitor mass of $\sim 260 \msun$, instead of $\sim 300 \msun$ as in all other cases. Interestingly, some cluster models (i.e. G1, G2, and N1) produce a small number of PISNe even at $Z=2\times 10^{-2}$, while isolated binaries do not. These differences reflect the fact that hardening can trigger additional interactions in systems that would otherwise remain detached.

\begin{figure*}[ht!]
\centering
\includegraphics[width=\textwidth]{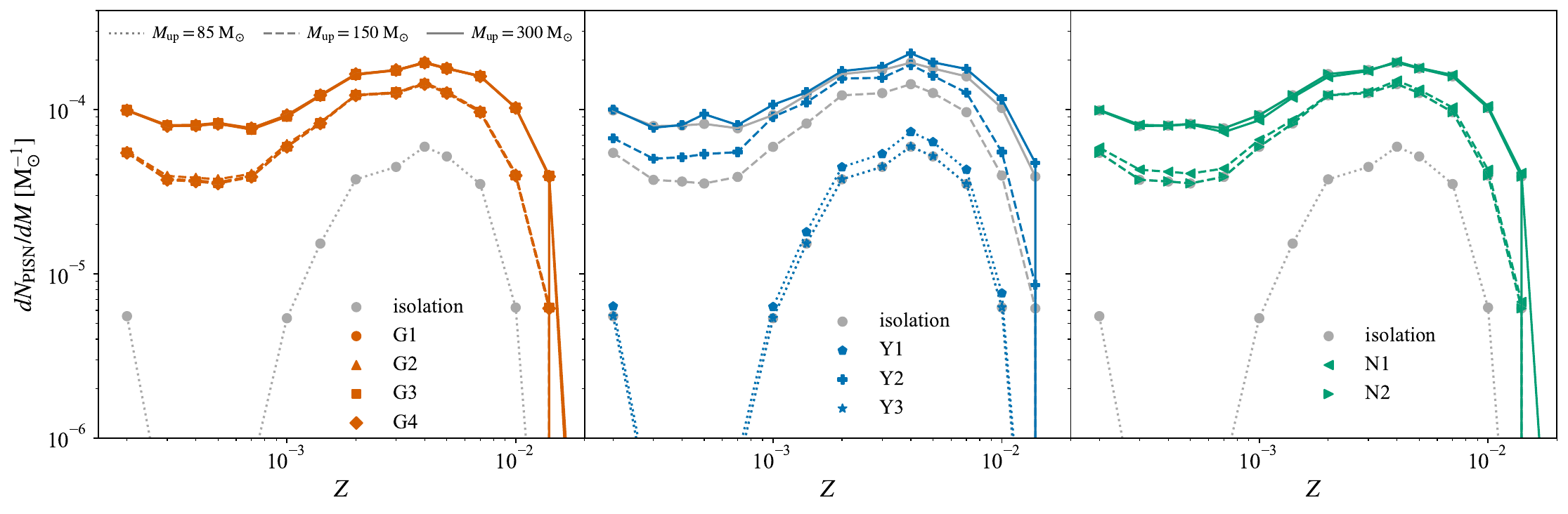}
\caption{PISN production efficiency for primordial binaries in dense star clusters. The three panels show GC-like (left), YSC-like (middle), and NSC-like (right) models. Solid and dashed lines indicate $M_{\rm up}=300\,\msun$ and $150\,\msun$, respectively. The $M_{\rm up}=85\,\msun$ case (dotted line) is shown only for Y1 and Y3, for which this upper mass limit is adopted as an exploratory low-mass-cluster scenario. Grey lines show the isolated-binary results for the corresponding values of $M_{\rm up}$, used as reference.}
\label{fig:dn_dm_hard}
\end{figure*}

Fig.~\ref{fig:dn_dm_hard} shows the PISN production efficiency for primordial binaries in all cluster models. The main result is that most clusters closely follow the isolated-binary case, as one can appreciate by the fact that most curves are superimposed on the grey ones. Binary hardening can shrink eccentric, initially wide binaries and make them interact, but such systems represent only a small fraction of the overall population of systems in the adopted primordial-binary initial conditions. Thus, the total PISN efficiency remains largely unchanged for most GC-like and NSC-like models. This behaviour is illustrated by the comparison between isolated binaries and cluster G2 in Appendix~\ref{sec:app_initial_properties}.

The largest deviation from isolation occurs for the most compact YSC-like model, Y2, which has the highest hardening rate in our cluster sample. In this case, hardening drives a larger fraction of binaries to interact, mainly through stellar mergers. This increases the PISN production efficiency by up to a factor $\lesssim2$ relative to isolated binaries. The enhancement is strongest for $M_{\rm up}=150\,\msun$, because in this case single-star-like progenitors above $150\,\msun$ are removed, while lower-mass binaries can still enter the PISN regime through mergers and mass transfer.

The same mechanism also suppresses part of the high-mass binary progenitor population in Y2. Very massive binaries that are forced to merge can produce remnants above the PISN mass range, which then collapse directly into IMBHs rather than exploding as PISNe. Thus, hardening creates new PISN progenitors from lower-mass binaries, but it can also remove very massive systems from the PISN channel. In Y2, the first effect dominates because lower-mass binaries are much more numerous.

A few cluster models with $M_{\rm up}=300\,\msun$ produce PISNe even at $Z=2\times10^{-2}$. These events are rare and come from extreme binaries with component masses close to the IMF upper limit and high eccentricities. Early collisions or mergers remove enough mass to bring the remnant into the PISN regime. In all other cases, the maximum metallicity for PISN production remains $Z=1.4\times10^{-2}$.

\subsection{Dynamical binaries}\label{sec:pisncluster_results_dyn}

\begin{figure*}[ht!]
\centering
\includegraphics[width=\textwidth]{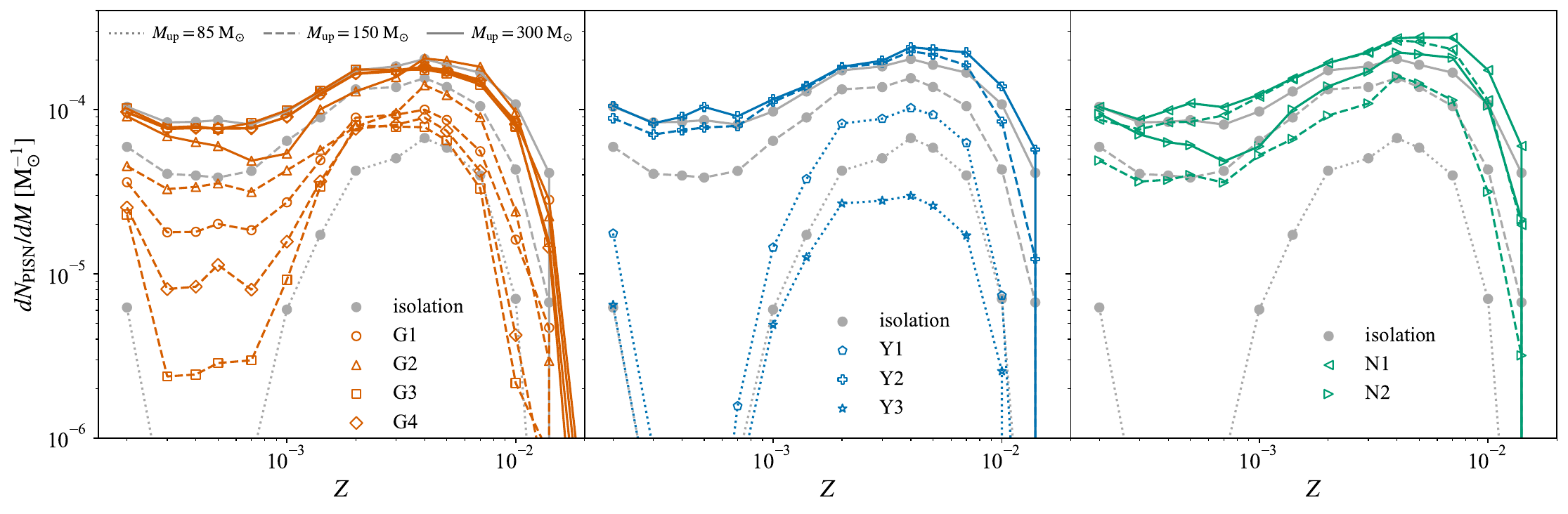}
\caption{Same as Fig.~\ref{fig:dn_dm_hard}, but for dynamical binaries.}
\label{fig:dn_dm_hard_dyn}
\end{figure*}

Fig.~\ref{fig:dn_dm_hard_dyn} shows the PISN production efficiency for the dynamically motivated binary initial conditions. Compared to primordial binaries, these models produce a wide range of efficiencies across cluster models. In some clusters, the PISN efficiency increases relative to isolated binaries; in others, it decreases. Thus, the main effect of dynamically motivated initial conditions is not a uniform enhancement as it depends on the host-cluster properties. Table ~\ref{tab:mzams_intervals_cluster} already hints at such features, in that the PISN progenitor mass ranges appear to be somewhat more fluctuating with respect to primordial binaries.

This behaviour is driven mainly by the initial distribution of semi-major axes and eccentricities. For dynamical binaries, semi-major axes are drawn from half-Gaussian distributions centered in the hard-binary separation, $a_h\propto(M_1+M_2)/\sigma_c^2$, depending on $\sigma_c$. The peak values and ranges of the semi-major axis distributions can vary by more than one order of magnitude between clusters with different $\sigma_c$ (e.g., Figs.~\ref{fig:props_G1_dyn_B1}, \ref{fig:props_G2_dyn_B1} in Appendix~\ref{sec:app_initial_properties}). Therefore, clusters with high velocity dispersion produce initially tighter hard binaries, which interact more easily through mass transfer and/or mergers, whereas wider binaries are more likely to avoid strong interactions. This makes the PISN efficiency sensitive to $\sigma_c$, not only to the hardening rate.  

For eccentricities, dynamical binaries are drawn from a thermal distribution, $p(e)=2e$, which favours high eccentricities and extends up to $e=1$. This differs from the primordial-binary model, where eccentricities are limited to $e=0.9$ and are weighted toward lower values (Sec.~\ref{sec:methods_clusters}). High eccentricities reduce the pericenter distance and favour early interactions, especially mergers and collisions. The combined effect of $a$, $e$, and component masses explains the larger scatter seen in Fig.~\ref{fig:dn_dm_hard_dyn}.

The strongest enhancement occurs in cluster N1. While Y2 has the highest hardening rate, N1 has the largest velocity dispersion in our sample,  $\sigma_c=96.5\rm\:km\:s^{-1}$, and thus the smallest hard-binary separation. Its dynamically assembled binaries are initially compact, interact efficiently, and are more likely to merge into the PISN progenitor range. This is why N1 reaches the highest PISN efficiencies among the dynamical-binary models. 

The same mechanism can also suppress PISN production. If mergers involve very massive binaries, the merger products can exceed the PISN mass range and collapse directly into IMBHs. In contrast, if the initial binaries are too wide, interactions remain inefficient and the PISN efficiency can fall below the isolated-binary case. This explains why the dynamically motivated models span both higher and lower efficiencies than the primordial-binary models.

Finally, dynamical binaries increase the number of rare PISNe at the highest metallicity, $Z=2\times 10^{-2}$. These events require extreme component masses, between 200 and $300\,\msun$, and very eccentric orbits with $0.8\leq e\leq 1$. In primordial binaries they are typically of order of tens, while in the dynamical-binary models they can reach several hundreds of events. They remain rare, however, and do not dominate the overall PISN production efficiency.

It is also worth noticing that Y3 is the only model with a lower maximum metallicity, $Z_{\rm max}=10^{-2}$. This is because Y3 combines the low IMF upper limit adopted for our least massive YSC-like models, $M_{\rm up}=85\:\msun$, with a relatively low hardening rate, $\mathcal{R}_{\rm hard}=1.4\times 10^2\ \rm pc^{-1}\:Myr^{-1}$. The number of PISNe at $Z=1.4\times 10^{-2}$ is already very small in the isolated-binary case and for Y1 with the same $M_{\rm up}$. In Y3, the weaker hardening further suppress the interacting binary channel, and no PISNe are produced at this metallicity.

\subsection{Cosmic PISN rate}\label{sec:pisn_rate_clusters}

We now convert the metallicity-dependent PISN production efficiencies, $dN_{\rm PISN}(Z)/dM_{\star}$, into cosmic PISN rates. We follow the framework of \citealp{Boco_2021,Gabrielli_2024a}, convolving $dN_{\rm PISN}(Z)/dM_\star$ with a metallicity-dependent star formation rate density (SFRD). We adopt $\sigma_{\rm Z}=0.15$ for the dispersion of galaxy metallicities around the fundamental metallicity relation, that we adopt to connect the stellar mass, star formation rate, and metallicity of star-forming galaxies. The full rate model is described in Appendix \ref{sec:app_pisn_rate}. Since the same galaxy-evolution model is used for all stellar-evolution channels, differences between the curves in this section mostly reflect differences in PISN production efficiency.

\begin{figure*}[ht!]
\centering
\includegraphics[width=\textwidth]{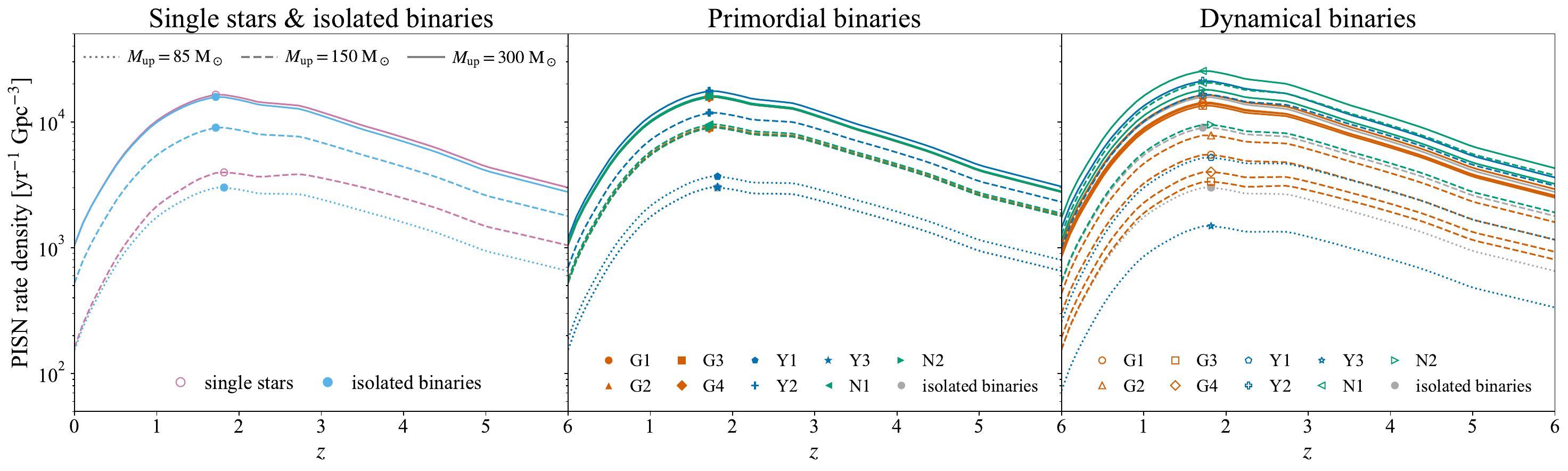}
\caption{Cosmic PISN rate as a function of redshift for the models explored in this work. The left panel compares single stars (pink lines) and isolated binaries (lightblue lines). The middle panel shows primordial binaries in dense star clusters, while the right panel shows dynamically-motivated binaries. Grey lines show the isolated-binary rates for comparison. Dotted, dashed and solid lines indicate the IMF upper limit: $M_{\rm up}=[85,\:150,\:300]\:\msun$, respectively. The $M_{\rm up}=85,\msun$ cluster case is shown only for Y1 and Y3. Markers indicate the redshift at which each rate peaks.}
\label{fig:pisn_rate_3panels}
\end{figure*}

Fig.~\ref{fig:pisn_rate_3panels} shows the resulting PISN rate as a function of redshift. The first panel (left) compares single stars and isolated binaries. The trends mirror those found for $dN_{\rm PISN}/dM_\star$ in Sec.~\ref{sec:pisncluster_results_star_binary}. In particular, for $M_{\rm up}=300\:\msun$, single stars and isolated binaries produce very similar rates, especially at $z<2$. For $M_{\rm up}=150\:\msun$, isolated binaries dominate by up to a factor three, depending on redshift, because binary interactions allow PISNe to form from lower initial masses. For $M_{\rm up}=85\:\msun$, single stars do not contribute, and the rate is entirely due to binaries. Interestingly, the latter contribution almost equals the solution for single stars with $M_{\rm up}=150\:\msun$ at $z<1$. Indeed, although the PISN production efficiency for isolated binaries with $M_{\rm up}=85\:\msun$ is lower (Fig.~\ref{fig:dn_dm_hard_singlebinary}), its higher $Z_{\rm max}$ provides a significant contribution to the PISN rate, especially at those low redshifts. Across these variations, the local rate spans from $\sim 1.5\times 10^2$ to $\gtrsim10^3\,{\rm yr^{-1}\,Gpc^{-3}}$, while the peak rate ranges from $\sim3\times10^3$ to $\sim1.5\times10^4\,{\rm yr^{-1}\,Gpc^{-3}}$. 

The second panel of Fig.~\ref{fig:pisn_rate_3panels} shows primordial binaries in dense star clusters. These rates are generally close to the isolated-binary case, especially for GC-like and NSC-like models with $M_{\rm up}=150$ and $300\:\msun$. This follows from Sec.~\ref{sec:pisncluster_results_ori_bin}, because for primordial binaries, hardening affects only a small fraction of initially wide and eccentric systems, so the integrated PISN rate does not change significantly. The main exception is Y2, whose high hardening rate increases the PISN rate by a factor $<2$, most clearly for $M_{\rm up}=150\:\msun$. The lowest rate comes from Y1 and Y3, when we adopt $M_{\rm up}=85\:\msun$, approaching the isolated-binaries solution.

The third panel of Fig.~\ref{fig:pisn_rate_3panels} shows the rates for dynamically motivated binaries. In this case, the scatter among cluster models is larger. At $z=0$, the PISN rate ranges from $\sim 70$ to $\sim 1.7\times 10^3\:\rm yr^{-1}\:Gpc^{-3}$, while the peak rate ranges from $\sim 1.4\times 10^3$ to $\sim 2.5\times 10^4\:\rm yr^{-1}\:Gpc^{-3}$. This larger spread comes from the strong dependence of the dynamical-binary initial conditions on cluster velocity dispersion, as discussed in Sec.~\ref{sec:pisncluster_results_dyn}. N1 gives the highest rate because its large velocity dispersion produces compact hard binaries that interact efficiently. Y3 gives one of the lowest rates because it combines $M_{\rm up}=85\:\msun$ with weak hardening. 

The redshift evolution is similar across all models, with the rate peaking at $z\lesssim 2$. This is expected because all channels are convolved with the same metallicity-dependent star formation history. It is worth noticing, however, that some lower models exhibit a slightly higher peak redshift, but still below $z=2$ (namely those with $M_{\rm up}=85\ \rm \msun$, single stars up to $150\ \rm\msun$, and roughly half of the dynamical cases). The reason is that, while generally $Z_{\rm max}=1.4\times 10^{-2}$, in the latter models we either find a lower $Z_{\rm max}=7\times 10^{-3}$, or a significantly suppressed PISN production efficiency at $Z_{\rm max}=1.4\times 10^{-2}$ (Sec.s~ \ref{sec:pisncluster_results_star_binary}, \ref{sec:pisncluster_results_ori_bin}, and \ref{sec:pisncluster_results_dyn}). Combined with the trend of our $Z$-dependent SFRD to decrease with redshift, the PISN rate tends to spread to higher redshifts. We refer the interested reader to \citealp{Gabrielli_2024a} for a detailed description of this interplay. The main effect of binary evolution, cluster hardening, and the IMF upper limit is therefore to change the normalization of the PISN rate, while the redshift at which it peaks is only slightly affected.

Overall, our models span more than one order of magnitude in cosmic PISN rate. The lowest rates are obtained for the low-$M_{\rm up}$ YSC-like models, especially Y3, while the highest rates are produced by isolated binaries with large $M_{\rm up}$, and by the most efficient cluster models, such as Y2 for primordial binaries, and N1 for dynamically motivated binaries.

\section{Discussion}\label{sec:discussion}

\subsection{The cosmic PISN rate}\label{sec:pisn_detection}

The framework presented in this work allows us to connect the metallicity-dependent PISN production efficiencies with the cosmic star-formation and metallicity history. Here, we combine the single-star, isolated-binary, and cluster-binary channels to estimate the total intrinsic cosmic PISN rate. For the cluster contribution, we use the two YSC-like models Y2 and Y3 with dynamically motivated initial conditions. These models bracket the range between our most efficient and least efficient cluster cases, and provide a simple estimate of the uncertainty associated with cluster properties. We do not include GC-like and NSC-like models in this combined rate, since they are expected to represent a smaller fraction of the total stellar mass in typical galaxies.

We weight the different channels using the binary fraction, $f_{\rm bin}$, and the cluster formation efficiency (CFE), $f_{\rm cfe}$. The fractions assigned to single stars, isolated binaries, and cluster binaries are $(1-f_{\rm bin})$, $f_{\rm bin}\cdot (1-f_{\rm cfe})$, and $f_{\rm bin}\cdot f_{\rm cfe}$, respectively. We consider two values for the binary fraction, $f_{\rm bin}=0.4$ and 1, spanning a broad range of plausible massive-star multiplicities \citep{Sana_2012,Moe_2017,Winters_2019}. We fix $f_{\rm cfe}=0.3$, representative of the upper range expected for YSCs \citep[e.g.][]{Mapelli_2021b}. 

\begin{figure*}[ht!]
\centering
\begin{subfigure}{\textwidth}
\centering
\includegraphics[width=0.8\textwidth]{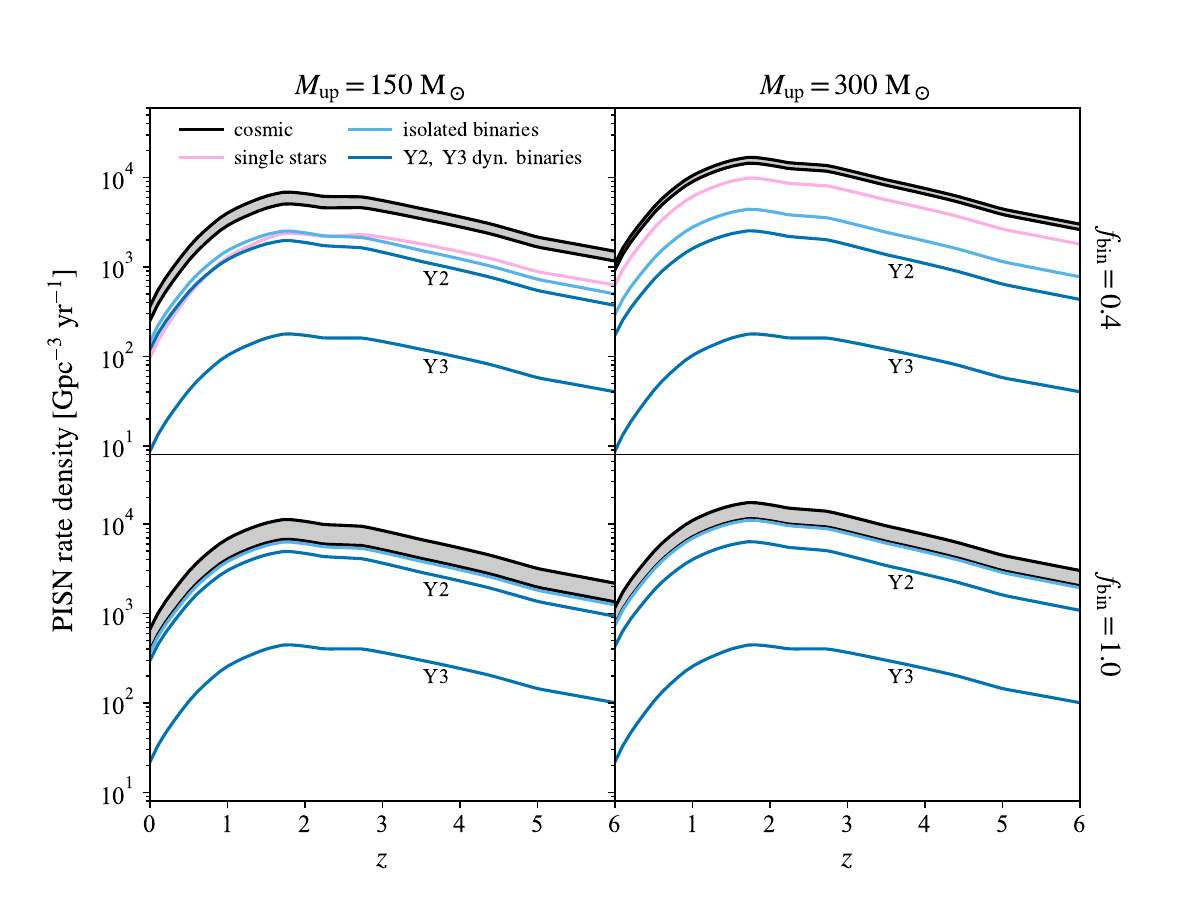}
\end{subfigure}
\caption{Intrinsic PISN rate density as a function of redshift for the combined single-star, isolated-binary, and cluster-binary channels. Pink, light-blue, and blue lines show the contributions from single stars, isolated binaries, and cluster binaries, respectively. For the cluster channel, we show the two YSC-like models Y2 and Y3 with dynamically motivated initial conditions, which bracket the range between our most efficient and least efficient cluster cases. Black lines show the total rate, with the grey band indicating the range obtained by using either Y2 or Y3 for the cluster contribution. The four panels show different combinations of $M_{\rm up}$ and $f_{\rm bin}$. The single-star contribution is absent when $f_{\rm bin}=1$.}
\label{fig:PISN_rate_channels}
\end{figure*}

Fig.~\ref{fig:PISN_rate_channels} shows the resulting intrinsic PISN rate density for different combinations of $M_{\rm up}$ and $f_{\rm bin}$. The total rate depends more strongly on $M_{\rm up}$ than on the assumed binary fraction. Increasing $M_{\rm up}$ from 150 to $300\:\msun$ can roughly triple the total PISN rate. Bringing $f_{\rm bin}$ from 0.4 to 1 suppresses the contribution from single stars, and boosts the binary channels, but the total rate changes by less than a factor of 2. The difference between using Y2 or Y3 for the cluster-binary channel is also less than a factor 2.

\begin{figure}[ht!]
\centering
\includegraphics[width=\columnwidth]{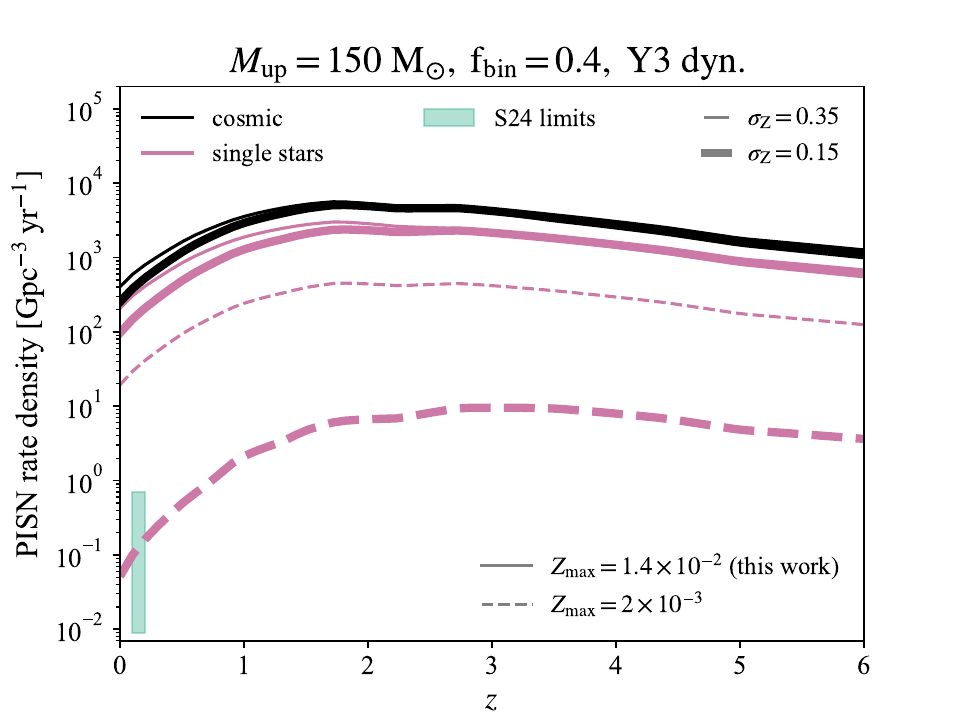}
\caption{Comparison between the PISN rate predicted with the \parsecii tracks used in this work and the rate obtained with the lower-$Z_{\rm max}$ \parseci tracks. We consider $M_{\rm up}=150\ \msun$, $f_{\rm bin}=0.4$, and Y3 dynamical binaries for the cluster contribution. Solid lines show the \parsecii case: the single-star contribution and the lower bound of the total cosmic rate from Fig.~\ref{fig:PISN_rate_channels}. Dashed lines show the single-star \parseci rate, for which $Z_{\rm max}=2\times 10^{-3}$, Thin and thick lines correspond to $\sigma_{\rm Z}=0.35$ and 0.15, respectively. The green band shows the PISN-rate limits inferred by \cite{Schulze_2024}}
\label{fig:PISN_rate_pessimistic}
\end{figure}

We now focus on the lowest-rate case, with $M_{\rm up}=150\:\msun$, $f_{\rm bin}=0.4$, and Y3 dynamical binaries as representative of the cluster population. Fig.~ \ref{fig:PISN_rate_pessimistic} compares this rate with the observational limits inferred by \citealp{Schulze_2024} from the PISN candidate SN~2018ibb, where we also consider an additional variation with $\sigma_{\rm Z}=0.35$, higher than the value of 0.15 adopted in this work. With the \parsecii tracks used in this work, the predicted rate lies between two and four orders of magnitudes above the limits by \citealp{Schulze_2024}. This is mainly driven by the high maximum metallicity of PISN progenitors in most of our models, $Z_{\rm max}=1.4\times 10^{-2}$. Since our metallicity-dependent SFRD peaks around $10^{-2}$, allowing PISNe up to near-solar metallicity gives a large contribution to the low-redshift rate. This confirms that the cosmic PISN rate is highly sensitive to the combination of $Z_{\rm max}$ and the galaxy metallicity dispersion, $\sigma_{\rm Z}$ \citep{Gabrielli_2024a}.

To explore the impact of a lower $Z_{\rm max}$, we also consider the \parseci stellar tracks \citep{Bressan_2012, Chen_2014, Chen_2015, Tang_2014}, corresponding to the version used in \citealp{Spera_2017} within \sevn (see also \citep{Gabrielli_2024a}).  With $M_{\rm up}=150\ \msun$ and the same PISN helium-core mass criterion adopted in this work, these tracks give $Z_{\rm max}=2\times 10^{-3}$, much lower than for \parsecii. For $\sigma_{\rm Z}=0.35$, this reduces the PISN rate by about one order of magnitude with respect to the \parsecii single-star case, but the rate still remains above the limits from \citealp{Schulze_2024}. However, if we also reduce the metallicity dispersion to $\sigma_{\rm Z}=0.15$, the \parsecii PISN rate gets suppressed by more than two additional orders of magnitude, becoming consistent with the \citealp{Schulze_2024} limits at $0.1<z<0.2$\footnote{We compute the PISN production efficiency for the \parseci tracks following \cite{Gabrielli_2024a}. This comparison includes only single-star progenitors and is meant to isolate the effect of lowering $Z_{\rm max}$.}. 

This comparison shows that current non-detections, or the rate inferred from candidates such as SN~2018ibb if confirmed as PISNe, can already constrain uncertain ingredients of stellar and galaxy evolution. In particular, lower values of $Z_{\rm max}$, around $2\times 10^{-3}$, and metallicity dispersions $\sigma_{\rm Z}\sim 0.15$, are more compatible with the \cite{Schulze_2024} limits. Such a low $Z_{\rm max}$ could result from stronger VMS winds \citep{Shepherd_2025,Simonato_2025}, stellar rotation \citep{Limongi_2018,Hirschi_2025}, or both. A detailed prediction of PISN detectability requires survey-specific selection functions, cadence, sky coverage, and detection efficiencies; we leave this to future work.

\subsection{Multiplicity of pair-instability supernovae}\label{sec:multiplicity}

Some binary systems produce one PISN, while others produce two. In our simulations, double-PISN systems occupy a restricted region of the initial parameter space. They require both components to be massive enough to enter the PISN regime, with initial masses comparable to those of single-star PISN progenitors (Table \ref{tab:mzams_intervals_cluster}). They also typically have semi-major axes above $\sim 10^3\:\rm R_{\odot}$, and eccentricities below 0.8 in most cases, depending on metallicity. These conditions allow the two components to evolve almost independently, or to interact only weakly, so that both stars can retain final helium-core masses in the PISN range.

Most binary PISN systems, however, produce only one explosion. This is because strong binary interactions often prevent both components from remaining in the PISN regime. In up-to $\gtrsim 90\%$ of binaries in the PISN progenitor populations, mergers combine the two stars into a single PISN progenitor. Less frequently, in typically $<10\%$ of the PISN binary populations, mass transfer can move one component into the PISN range while moving the other out of it. Since double-PISN systems require two initially VMSs and relatively weak interactions, they only account for $<10\%$ of one-PISN binaries.

This distinction may be relevant for interpreting future PISN candidates. In our models, many one-PISN binaries originate from stellar mergers, as discussed in Sec.~\ref{sec:results} and Appendix~\ref{sec:app_initial_properties}. A merger origin has also been suggested for some PISN candidates, including SN~2023vbw, based on light-curve features consistent with interaction with a disk-like circumstellar medium \citep[see][]{Hiramatsu_2026}. This does not provide a direct identification of the progenitor channel, since other interpretations remain possible. However, a future sample of PISNe could help constrain how often PISN progenitors are shaped by binary interactions, and therefore provide indirect information on the multiplicity of VMSs.

We finally point out that, as shown in \cite{Mestichelli_2026}, in the case of binaries in young massive clusters (corresponding roughly to G1-G4; \citealp{Bastian_2013}), PISNe form primarily as single objects. In fact, we expect most PISNe in such clusters to derive from the merger of a primordial binary system, due to the combination of a high fraction of primordial binaries, hard semi-major axes, and an IMF upper limit of $150\,\msun$. Nevertheless, more compact clusters, or clusters with initial binary properties similar to the dynamical conditions explored in this paper, might be able to form double PISNe.

\subsection{Stellar mergers and phases}\label{sec:phases}

We now focus on binary PISN progenitors that undergo a stellar merger. The goal is to identify which merger channels dominate and whether the merger happens while the two stars are still on the MS or after significant stellar evolution. 

In \sevn, stellar mergers are classified according to the event that triggers them \citep{Iorio_2023}. A merger can be triggered by CE evolution  or RLO, giving ``CEMerger'' and ``RLOMerger'' events, respectively. It can also follow a direct stellar collision, classified as ``CollMerger''. We refer to mergers that are not associated with any of these events as ``Merger''. Multiple triggers are also possible.

Most merging binary PISN progenitors undergo either Merger or CollMerger events. For example, for isolated binaries with $M_{\rm up}=300\:\msun$ and $Z=2\times 10^{-3}$, about $83\%$ of PISN progenitors experience a Merger event, while $\sim 16\%$ undergo a CollMerger. In cluster Y2 (i.e. our most hardening-efficient cluster), for the same metallicity, IMF upper limit, and primordial-binary initial conditions, these fractions remain similar, $\sim 82\%$ and $\sim 18\%$. In contrast, with dynamically-motivated initial conditions in Y2, the CollMerger fraction increases to $\sim 68\%$, while the Merger fraction decreases to $\sim 32\%$. This is expected because dynamical binaries are initialized with high eccentricities, which reduce the pericentre distance and favor direct collisions. Other merger classes are rare; CEMerger is the most common among them, but still accounts for less than $1\%$ of binary PISN progenitors.

\begin{figure*}[ht!]
\centering
\includegraphics[width=\textwidth]{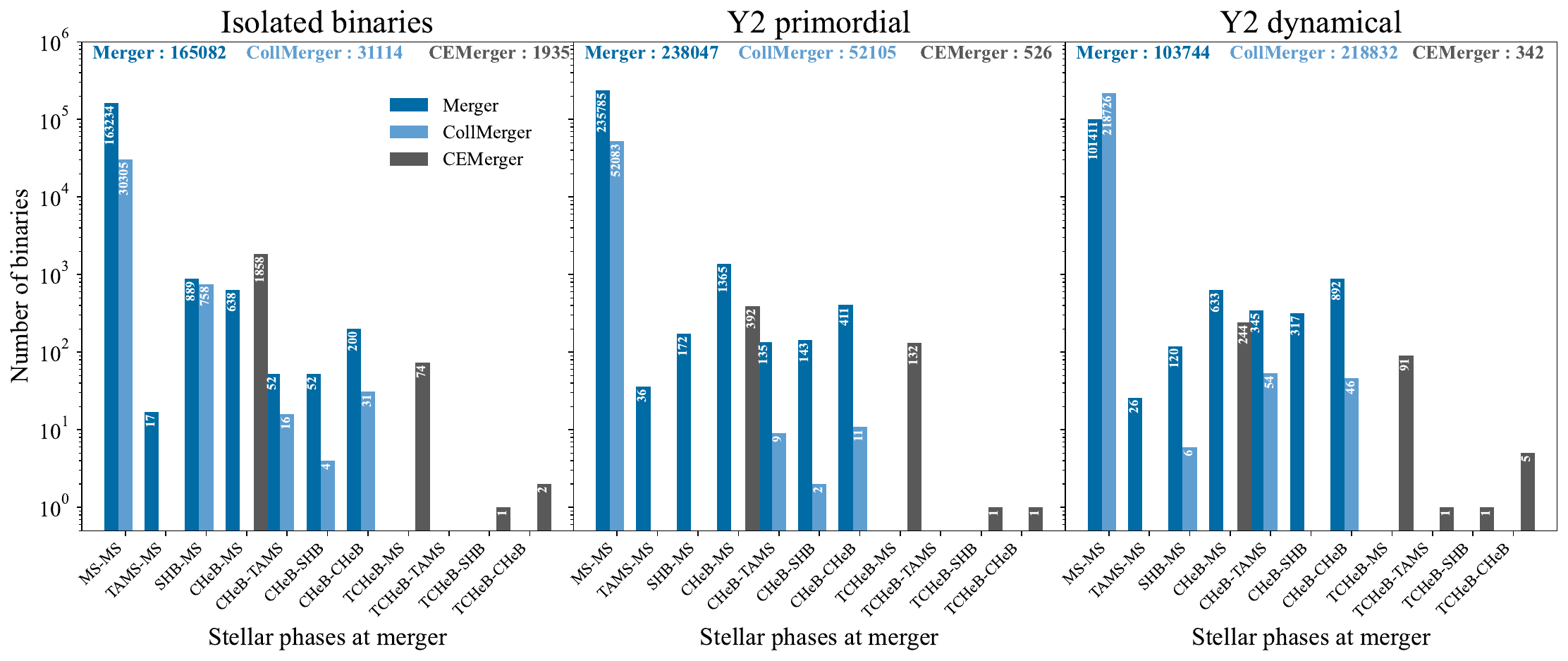}
\caption{Stellar phases of the binary components immediately before merger, for systems that undergo pure mergers, collision-induced mergers, and CE-induced mergers. Left, middle, and right panels show the cases for isolated binaries, primordial, and dynamical binaries in cluster Y2, respectively. $M_{\rm up}$ is fixed to $300\:\msun$, and $Z=2\times 10^{-3}$.}
\label{fig:phases}
\end{figure*}

Fig.~\ref{fig:phases} shows the stellar phases of the two components immediately before merger, for Merger, CollMerger, and CEMerger events. For Merger and CollMerger events, the vast majority of systems merge while both components are still on the MS. At $Z=2\times10^{-3}$ and $M_{\rm up}=300\,\msun$, evolved systems account for only $\sim2$--$3$\% of the Merger or CollMerger populations, both in isolation and in Y2, and for both primordial and dynamically motivated initial conditions.

In contrast, the CEMerger subpopulation involves at least one component that has already left the MS, as expected for mergers triggered by CE evolution. However, because CEMerger systems are rare, they do not affect the overall conclusion, which is that most merging binary PISN progenitors merge early, before either component undergoes advanced stellar evolution. This result is robust against changes in environment, metallicity, and binary initial conditions within the models explored here.

This has implications for the interpretation of merger-origin PISN candidates. For example, \cite{Hiramatsu_2026} proposed that SN~2023vbw may originate from a post-core-He burning merger. Such a channel is not dominant in our models, where most merging PISN progenitors merge while both stars are still on the MS. However, the progenitor channel of SN~2023vbw represents only one individual system, and it is not uniquely established, whereas our results describe the dominant channels in a population-synthesis framework.

\subsection{Caveats}\label{sec:discussion_other_interactions}

Our results should be interpreted in light of several caveats. First, we do not perform direct dynamical simulations of star clusters. We use \sevn to evolve large populations of binaries, with $10^7$ systems per metallicity and model variation, and include the secular effect of binary hardening through semi-analytic prescription. This approach is necessary because PISNe are rare, with a production efficiency down to order $10^{-6}\,\msun^{-1}$ or even lower. However, it does not follow individual few-body encounters, exchanges, ejections, binary--binary interactions, or physical collisions induced by close encounters. These missing processes can affect the evolution of massive binaries in dense clusters, especially for high binary fractions, $f_{\rm bin}>0.4$ \citep[e.g.,][]{Mikkola_1983,Zevin_2019,Arca_Sedda_2024}. We partly bracket their possible impact by considering both primordial binaries and dynamically-motivated hard binaries. Nonetheless, this is not equivalent to a self-consistent dynamical treatment. In real clusters, dynamically assembled binaries form after some delay, set by the relevant encounter and relaxation timescales. Our dynamically motivated models do not include stellar evolution during this delay. A direct, more self-consistent, dynamical approach including all the mentioned ingredients would be computationally prohibitive for the broad parameter space explored here.

Second, we adopt the central density and velocity dispersion of each cluster when computing the hardening rate. This choice is motivated by the expectation that massive stars and massive binaries preferentially reside in dense inner regions. However, the mass-segregation timescale can be comparable to, or longer than, the lifetime of the most massive stars. Binaries born or residing at larger radii would experience lower densities and weaker hardening. Our cluster calculations should therefore be regarded as an optimistic estimate of the impact of binary hardening.

Third, our hardening prescription captures only the secular shrinkage of hard binaries due to repeated binary--single encounters. It does not include strong impulsive encounters that can directly trigger collisions or mergers. Such events are expected to be more common in the densest cluster regions and could provide an additional channel for producing merger-origin PISNe. We leave their implementation to future work.

Fourth, our treatment of NSCs is simplified. NSCs can have complex formation histories, with multiple stellar populations, repeated star-formation episodes, and interactions with a central massive black hole \citep[e.g.,][]{Neumayer_2020}. Here we model them as single stellar populations with fixed structural properties. Our NSC models therefore capture only some global properties of nuclear clusters, e.g., their large masses, high densities, and high velocity dispersions, and they should not be interpreted as a complete model for the formation and evolution of NSCs.

Fifth, we adopt non-rotating stellar-evolution tracks. Rotation can modify the internal mixing, final helium-core mass, mass-loss history, and therefore the metallicity range of PISN progenitors \citep[e.g.,][]{Limongi_2018,Hirschi_2025}. Exploring the combined effect of rotation, binary evolution, and cluster dynamics is beyond the scope of this work.

Sixth, in all our SEVN simulations, we fix the value of $\alpha_{\rm CE}=3$ for the efficiency of the CE phase. We do not expect that changing this parameter would significantly alter our conclusions since, as shown e.g. in Sec.~ \ref{sec:phases} and Appendix \ref{sec:app_initial_properties}, we find that systems experiencing CE - and, more generally, mass transfer not resulting in merger - always represent a subdominant population.

Finally, as described in Sec.~ \ref{sec:methods}, we use the O abundance to define the gas-phase metallicity for our galaxy evolution model and cosmic PISN rate computation (Appendix \ref{sec:app_pisn_rate}). Recent works by \citealp{Chruslinska_2024,Chruslinska_2025} argue in favor of adopting the Fe abundance instead, that would more-accurately describe stellar evolution and feedback. It would be interesting to explore the effect of such a correction on our results, which we leave to future work.

\section{Conclusions}\label{sec:conclusions}

In this paper, we studied how binary evolution and dense stellar environments affect the production of PISNe. We used \sevn with \parsecii stellar-evolution tracks to evolve populations of single stars, isolated binaries, and binaries in dense star clusters. For cluster binaries, we explored both primordial-binary initial conditions and dynamically motivated hard-binary initial conditions, and we followed the secular effect of binary hardening. We then converted the resulting metallicity-dependent PISN production efficiencies into cosmic PISN rates. 

We find that binary evolution can substantially broaden the initial mass range of PISN progenitors. While single stars produce PISNe only for ZAMS masses above $\sim 110\,\msun$ in our models, isolated binaries can produce PISNe from components with initial masses down to $\sim 40\,\msun$. These lower-mass progenitors are mainly produced through stellar mergers and mass transfer. Binary evolution can also move stars out of the PISN regime, for example by increasing their mass above the PISN range and leading to direct collapse into IMBHs. 

The impact of binaries on the PISN production efficiency depends strongly on the IMF upper mass limit. For $M_{\rm up}=300\:\msun$, single stars are more efficient than isolated binaries by less than a factor of 2 at most metallicities. For $M_{\rm up}=150\:\msun$, instead, isolated binaries dominate because they can still produce PISNe through interactions from systems below the single-star PISN mass range. In this case, the cosmic PISN rate from isolated binaries can be up-to three times higher than the single-star contribution. For $M_{\rm up}=85\:\msun$, single stars do not produce PISNe in our models, while binaries can still contribute significantly through mergers and mass transfer. 

Binary hardening in dense star clusters does not produce a uniform effect. For primordial binaries, most cluster models remain close to the isolated-binary case, because only a small fraction of initially wide and eccentric binaries are strongly affected by hardening. The largest enhancement occurs in the compact YSC-like model Y2, where the PISN production efficiency increases by a factor $<2$. For dynamically motivated binaries, the scatter among cluster models is larger, spanning $\gtrsim 1$ order of magnitude. In this case, the outcome depends strongly on the cluster velocity dispersion, which sets the initial hard-binary separation. The NSC-like model N1 gives the highest PISN efficiencies, while Y3 gives some of the lowest values because it combines $M_{\rm up}=85\:\msun$ with weak hardening.

Across all channels and model variations, the cosmic PISN rate spans more than one order of magnitude. The redshift evolution is similar, with the rate peaking at $z\lesssim 2$, since most channels yield the same $Z_{\rm max}=1.4\times 10^{-2}$, and they are all convolved with a fixed metallicity-dependent star-formation history. The main differences between models are therefore in the normalization of the rate, driven by the IMF upper limit, binary interactions, and the assumed cluster environment.

The framework presented here connects metallicity-dependent PISN production efficiencies with the cosmic star-formation and metallicity history, accounting for single stars, isolated binaries, and binaries in dense stellar environments. Building on these results, in a future work we will delve into the observability of PISNe, and predict their detection rate with current and future facilities. Several works predict the achievement of the first confident PISN observations in the near future, with instruments such as ZTF, JWST, Rubin, Euclid, Roman, and ULTIMATE-Subaru (e.g. \citealt{Weinmann_2005,Whalen_2012,Smidt_2015,Kozyreva_2014,Hartwig_2018,Regos_2020,Moriya_2019,Moriya_2022,Moriya_2022_1,Tanikawa_2022,Tanikawa_2024,Venditti_2024,Jeon_2026}). Some particularly promising candidates have already been proposed, including SN~2018ibb (\citealp{Schulze_2024}) and SN~2023vbw (\citealp{Hiramatsu_2026}). Moreover, considerable efforts are being dedicated to developing optimal observational strategies, that would enable to identify PISN candidates, and distinguish them from other transient types (\citealp{Moriya_2022,Moriya_2022_1,Moriya_2023,Moriya_2025,DeCoursey_2025}). The comparison of PISN detections with our results on the cosmic PISN rate, would enable key constraints on several uncertain aspects of VMS formation and evolution. These comprise the IMF upper limit, the onset physics of PISNe, and the maximum metallicity of PISN progenitors, serving as a guide for stellar evolution codes. Furthermore, it would constrain uncertain quantities in galaxy evolution models, including the dispersion of galaxy metallicities across cosmic time. Interestingly, assuming that SN~2018ibb was an actual PISN (\citealp{Schulze_2024}), might already point towards $Z_{\rm max}\sim 2\times 10^{-3}$, due to e.g. enhanced stellar-wind mass loss and/or rotation, and $\sigma_{\rm Z}$ around 0.15.

\section*{Acknowledgements}
We thank Ragnhild Lunnan for helpful discussions.
FG and EZ acknowledge financial support from the Carl Trygger Foundation for scientific research (grant CTS 24: 3297). EZ acknowledges funding from project grant 2022-03804 from the Swedish Research Council (Vetenskapsr\aa{}det) and grant 2025-00213 from the Swedish National Space Agency.
MS acknowledges support from the INAF-Large Grant 2024:”Envisioning Tomorrow: prospects and challenges for multimessenger astronomy in the era of Rubin and Einstein Telescope”, from Fondazione ICSC, Spoke 3 Astrophysics and Cosmos Observations, National Recovery and Resilience Plan (Piano Nazionale di Ripresa e Resilienza, PNRR) Project ID CN\_00000013 “Italian Research Center on High-Performance Computing, Big Data and Quantum Computing” funded by MUR Missione 4 Componente 2 Investimento 1.4: Potenziamento strutture di ricerca e creazione di “campioni nazionali di R$\&$S (M4C2-19 )” - Next Generation EU (NGEU), and from the program “Data Science methods for Multi-Messenger Astrophysics $\&$ Multi-Survey Cosmology” funded by the Italian Ministry of University and Research, Programmazione triennale 2021/2023 (DM n.2503 dd. 09/12/2019), Programma Congiunto Scuole.
MAS acknowledges funding from the European Union’s Horizon 2020 research and innovation program under the Marie Skłodowska-Curie grant agreement No.~101025436 (project GRACE-BH) and from the MERAC Foundation. 
GC acknowledges financial support from the European Union–Next Generation EU, Mission 4, Component 2, CUP: C93C24004920006, project ‘FIRES’.
This paper is supported by the European Union’s Horizon Europe research and innovation programme under grant agreement No 101131928, project ACME.

\bibliography{pisne_in_cluster}{}
\bibliographystyle{aa}

\appendix

\section{Initial properties of binary PISN progenitors}\label{sec:app_initial_properties}

In this section, we show the distributions of initial properties for binary PISN progenitors, namely their initial component masses, $M_1$ and $M_2$, semi-major axis, $a$, and eccentricity, $e$. This is useful to understand the reason of the $dN_{\rm PISN}/dM_{\star}$ dependence on stellar environment, i.e. formation channel, cluster properties, IMF upper limit, and metallicity, described in Sec.~\ref{sec:pisncluster_results_ori_bin}.

\begin{figure*}[ht!]
\centering
\includegraphics[scale=0.7]{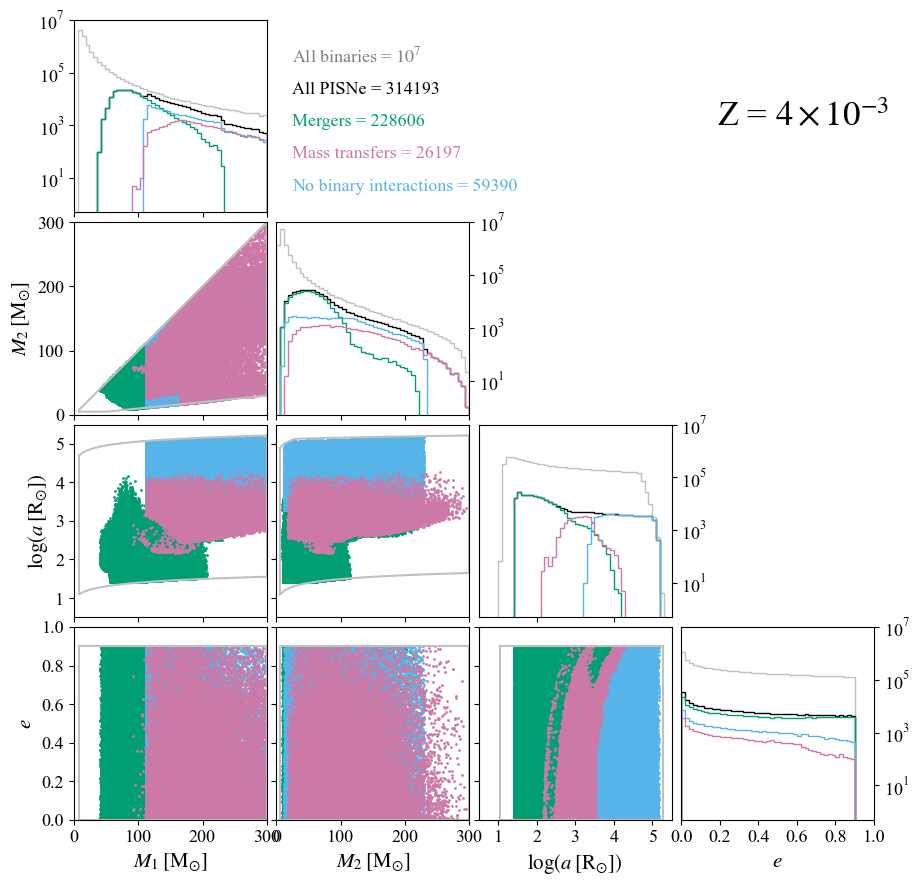}
\caption{Initial distributions of ZAMS component masses, semi-major axis and eccentricities of PISN progenitors, for the case of isolated binaries. Metallicity is fixed to $Z=4\times 10^{-3}$, and $M_{\rm up}=300\:\msun$. We distinguish between binaries that merge (indicated in green), binaries that do not merge but experience mass-transfer (e.g. RLO or CE, in pink), and binaries that do not interact at all (blue). We also show the total PISN contribution in black, and the whole underlying binary population in grey (for the latter, we only show the distribution boundaries). The number of events in each category is also reported, in corresponding colors.}
\label{fig:props_nohard}
\end{figure*}

\begin{figure*}[ht!]
\centering
\includegraphics[scale=0.7]{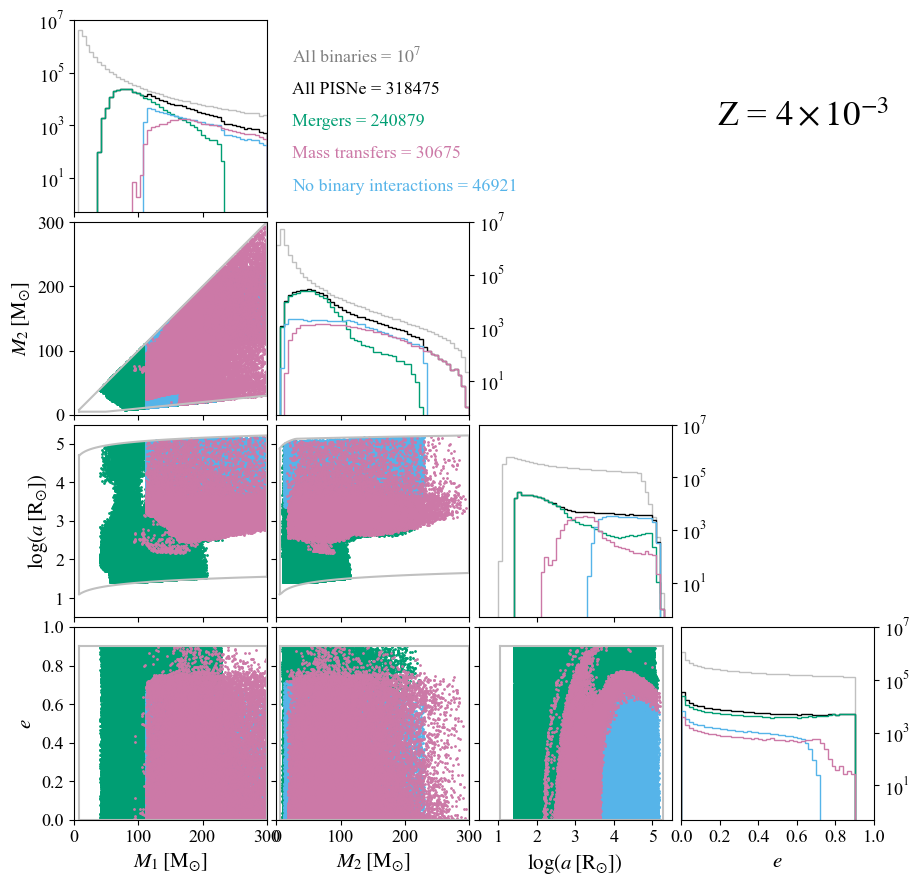}
\caption{Same as Fig.~\ref{fig:props_nohard}, for cluster G2.}
\label{fig:props_G2}
\end{figure*}

Fig.s~\ref{fig:props_nohard} and \ref{fig:props_G2} show the cases of isolated-binaries and cluster G2 with initial conditions for primordial binaries, respectively. We present results for $Z=4\times 10^{-3}$, i.e. the peak-metallicity of the PISN efficiency, and $M_{\rm up}=300\:\msun$. Different colors indicate binaries that experience mergers (green), mass-transfer events, either stable or unstable (mainly RLO and CE, pink), and those that do not interact at all (light blue). The total PISN distributions, comprising all of these cases, are displayed in black. Grey lines show the boundaries of the whole simulated binary population, for comparison. We also report the number of binaries in each subpopulation, in corresponding colors.

We see that cluster G2 produces slightly more PISNe with respect to isolation ($\sim 3.18\times 10^5$ against $3.14\times 10^5$). In particular, G2 slightly enhances the number of mergers and mass-transfer events, and decreases the number of binaries that do not interact. As anticipated in Sec.~\ref{sec:results}, the reason for these differences lies among those binaries with high semi-major axis and eccentricity. As can be seen e.g. by comparing the $\log(a)-e$ panels, while in isolation such binaries are not interacting, hardening in G2 induces them to merge, or interact in other ways. Indeed, the effect of hardening is to reduce the binary separation and increase its eccentricity, favoring interactions between the components. On a more general level, the three subpopulations identify somewhat different semi-major axis regions. While merging binaries constitute the bulk of the PISN progenitor population at low semi-major axis, $1.5\lesssim\log(a)\lesssim 2.5$ (despite a tail at higher values), non-interacting ones only exhibit high semi-major axes between $\log(a)\sim 3.5$ and $\log(a)\gtrsim 5$. Binaries that experience mass transfer tend to lie in between the two. The reason for this behavior is that systems with smaller separations favor interactions and mergers, while at increasing semi-major axis this becomes more and more difficult. First binaries only interact without merging, and then they are simply too wide to have any sort of interaction. On the other hand, the distributions of eccentricities are all analogous in shape, only differing in the upper end. Indeed, the higher the eccentricity, the closer the components will get, leading them to exchange mass or even merge. These high-semi-major axis, high-eccentricity binaries, only represent a small subpopulation. Therefore, binary hardening only has a secondary effect on the overall population, that is instead dominated by systems with small semi-major axes and eccentricities. This explains why their PISN production efficiencies are so similar.

The distributions of component masses are more or less unvaried in the isolated and G2 cases, except for a slight change in normalization due to the effect just described. As one can see from the $M_1-M_2$ panel and the corresponding marginal distributions, the primary masses of non-interacting binaries lie above $112\:\msun$, i.e. the minimum mass of single-star PISN progenitors according to the adopted stellar evolution model (Table \ref{tab:mzams_intervals_cluster}). Indeed, due to the absence of interactions, these binary components behave exactly as if they were single stars. For binaries experiencing mass transfer, the $M_1$ distribution starts from slightly lower values, since processes like RLO or CE can lead also lower-mass stars to explode as PISNe. Furthermore, these two subpopulations provide PISN progenitors with $M_1$ up to $\sim 300\:\msun$, higher than $230\:\msun$ for single stars (Table \ref{tab:mzams_intervals_cluster}). On one hand, stars that are initially too massive to produce PISNe, can lose material via mass transfer, and enter the PISN regime. Moreover, while in non-interacting binaries primary components with mass between $230$ and $300\:\msun$ do not produce PISNe themselves, they can witness the PISN explosion of their companion.

However, the processes that are most effective in producing PISNe in binaries, are stellar mergers. Indeed, they extend the primary masses of PISN progenitors to much lower values than the single-star case, down to $40\:\msun$ (for primary stars). We find that merging binaries represent the bulk of the PISN progenitor population, accounting for almost 3 times the number of interacting and non-interacting binaries combined. On the other hand, all $M_2$ distributions start from masses close to the lower limit of the initial population. These are binaries where only the primary component explodes as PISNe, while the secondary can have any given mass (see also Sec.~\ref{sec:multiplicity}). The tail at high-masses of the $M_2$ distribution is composed only of systems that experience mass transfer. As we show again in Sec.~\ref{sec:multiplicity}, these are mostly systems where both primary and secondary explode as PISNe, due to their extreme masses. Merging systems only exhibit maximum primary(secondary) masses of $\lesssim 250\:\msun$($\gtrsim 200\:\msun$). This is because binaries with higher component masses merge into stars that are too massive to produce PISNe, and instead directly collapse into IMBHs. These binaries are thus removed from the PISN progenitor population. Finally, the fact that all mass distributions tend to decrease going to higher values, is due to the trend of the IMF, as can be seen from the grey distributions.

Reducing $M_{\rm up}$ to $150\:\msun$ completely removes the PISN contribution coming from higher-mass progenitors. Below $150\:\msun$, the mass distributions are identical to the $M_{\rm up}=300\:\msun$ case. As a consequence, the semi-major axis and eccentricity distributions are suppressed in normalization, while maintaining the same overall shape. These effects are somewhat enhanced for $Z<10^{-3}$ and $Z\geq10^{-2}$, since PISN progenitors tend to distribute to higher masses with respect to other metallicities. This is an effect of both stellar evolution model, and binary evolution prescriptions, allowing merger remnants at low metallicity to produce PISNe starting from higher masses, with respect to higher metallicity.

\begin{figure*}[ht!]
\centering
\includegraphics[scale=0.7]{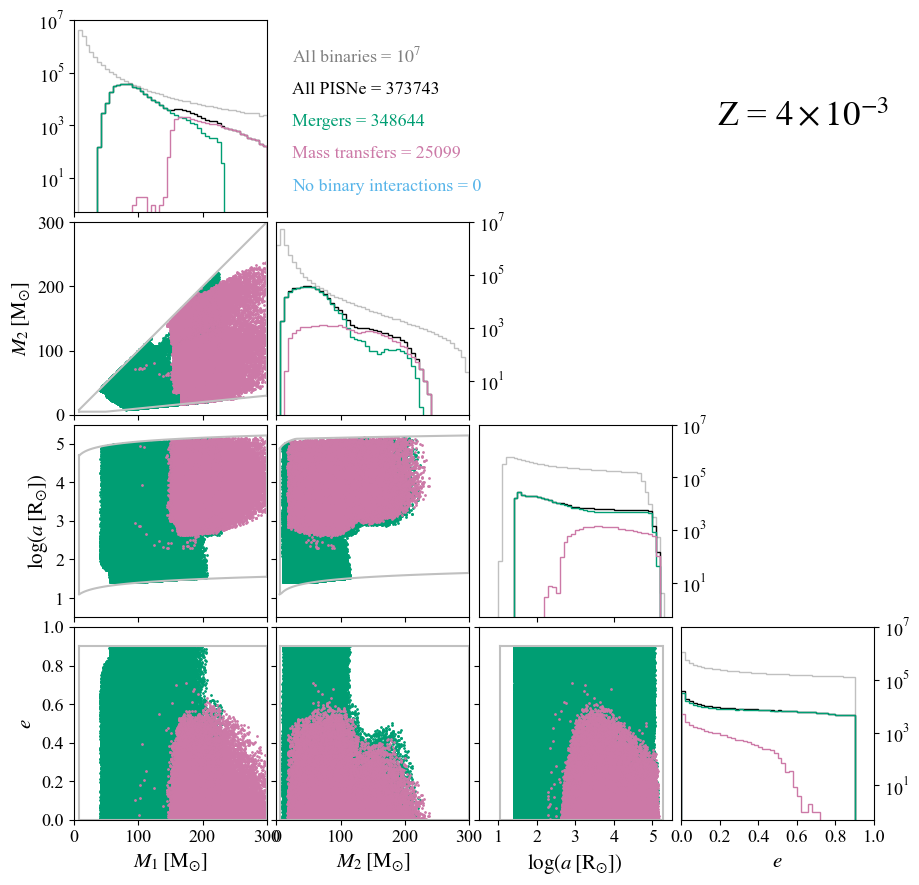}
\caption{Same as Fig.~\ref{fig:props_G2}, for cluster Y2.}
\label{fig:props_Y2}
\end{figure*}

Differently from our GCs, Y2 exhibits both high $\rho_c$ ($6.6\times 10^6\:\msun\:\rm pc^{-3}$) and low $\sigma_c$ ($2.2\:\rm km\:s^{-1}$), enhancing the hardening rate (Table \ref{tab:clusters}). As can be seen in Fig.~\ref{fig:props_Y2} for $M_{\rm up}=300\:\msun$ and $Z=4\times 10^{-3}$, the effect on the population of binary PISN progenitors is significant. The number of non-interacting binaries goes to zero, with the region at high-$a$ and low-$e$ populated by mass-transfer systems. The rest of the $a-e$ plane is occupied by merging binaries, that increase in number with respect to both isolation and G2 cases. The total number of PISNe increases from $\sim 3.14\times 10^5$ for the case of isolated binaries, and $\sim 3.18\times 10^5$ for G2, to $\sim 3.73\times 10^5$. The primary distribution for mass-transfer binaries starts at higher values than the previous cases, around $150\:\msun$ (with some exception at $\lesssim 100\:\msun$). Imposing an IMF upper limit of $M_{\rm up}=150\:\msun$, removes almost completely the mass-transfer subpopulation. However, the bulk of merging PISN progenitors is still present, providing a higher PISN efficiency than the $M_{\rm up}=150\:\msun$, GC cases.

\begin{figure*}[ht!]
\centering
\includegraphics[scale=0.7]{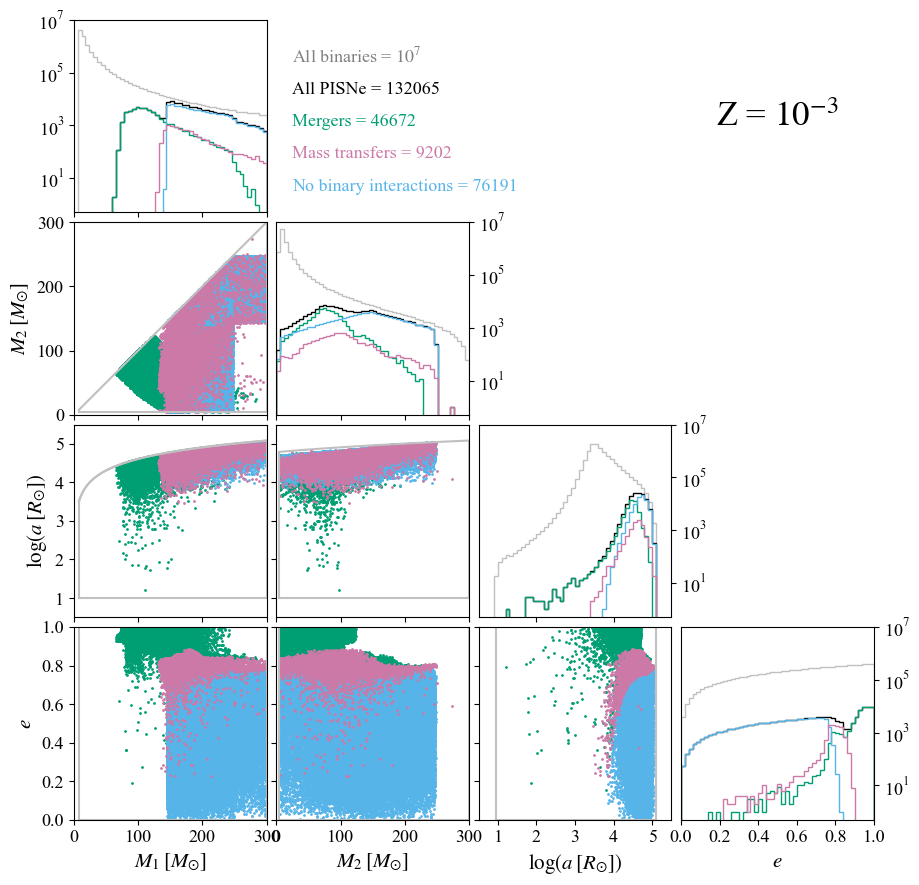}
\caption{Same as Fig.~\ref{fig:props_G2}, for cluster G1 with dynamical binaries, and $Z=10^{-3}$.}
\label{fig:props_G1_dyn_B1}
\end{figure*}

\begin{figure*}[ht!]
\centering
\includegraphics[scale=0.7]{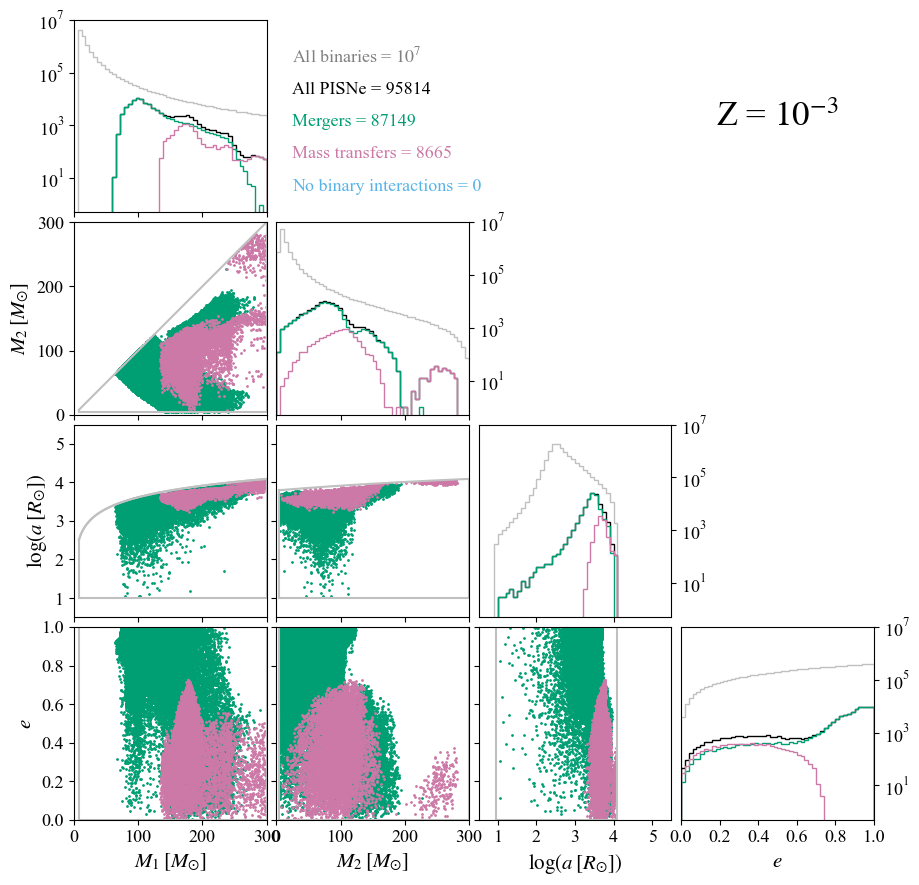}
\caption{Same as Fig.~\ref{fig:props_G2}, for dynamical binaries and $Z=10^{-3}$.}
\label{fig:props_G2_dyn_B1}
\end{figure*}

We now turn our attention to the effect of varying initial conditions for the binary population, from those describing primordial binaries, to those representative of dynamical binaries. Fig.s~\ref{fig:props_G1_dyn_B1}, \ref{fig:props_G2_dyn_B1} show the progenitor property distributions for clusters G1 and G2 with dynamical initial conditions, respectively, for $Z=10^{-3}$ (where the difference in PISN production efficiency between these two clusters is higher). The initial semi-major axis distributions are shifted by 1 order of magnitude ($\log(a)\sim 1$) between clusters G1 and G2, due to their dependence on $\sigma_c$ highlighted above. Since G1 has lower $\sigma_c$ than G2 ($9.7\:\rm km\:s^{-1}$ against $30.5\:\rm km\:s^{-1}$), semi-major axes distribute at higher values for the former cluster. In particular, G1 also covers the range $4<\log(a)<5$, that provides all PISNe from systems with no binary interactions (light-blue distributions in Fig.s~\ref{fig:props_G1_dyn_B1}, \ref{fig:props_G2_dyn_B1}). G2 only reaches $\log(a)\lesssim 4$, and thus does not contain non-interacting PISN progenitors. Mass-transfer systems are slightly favored in G1, while G2 produces more PISNe from mergers, due to the higher hardening rate. Overall, the difference in non-interacting systems prevails, and the total number of PISNe turns out to be greater for G1 than G2 ($\sim 1.32\times 10^5$ against $\sim 9.6\times 10^4$). It is also to be noted that, while in most cases stellar mergers are the main drivers of PISN production, for dynamical clusters with low $\sigma_c$ such as G1 it is the non-interacting systems that dominate, due to the higher semi-major axes.

These examples show how the dynamical initial conditions can affect in a non-trivial way the PISN production efficiency and the properties of PISN progenitors, due to the interplay between the initial semi-major axis, eccentricity, and component-mass distributions.

\section{Cosmic PISN rate computation}\label{sec:app_pisn_rate}

In this section, we show how we convert the PISN production efficiency, $dN_{\rm PISN}/dM_{\star}(Z)$, defined in Sec.~\ref{sec:methods_pisn_efficiency}, to the cosmic PISN rate. 

Following \cite{Gabrielli_2024a}, we convolve $dN_{\rm PISN}(Z)/dM_{\star}$ with a $Z$-dependent SFRD, $d^3 M_{\star}(Z,z)/dt\,dV\,d\log Z$, that quantifies the amount of star-forming mass available at a given redshift and metallicity, per unit time, comoving volume, and metallicity \citep[e.g.,][]{Madau_2014,Chruslinska_2019,Chruslinska_2021,Boco_2021}:
\begin{equation}
\frac{d^2 N_{\rm PISN}}{dt\,dV}(z)=\int d\log Z\,\,\frac{dN_{\rm PISN}}{dM_{\star}}(Z)\times\frac{d^3 M_{\star}}{dt\,dV\,d\log Z}(Z,z).
\end{equation}
We note that here we do not consider any delay time between binary formation and PISN explosion, since we find that in most cases the evolution times of PISN binary progenitors, $t_{\rm ev}$, are below $5\:\rm Myr$. As in \cite{Gabrielli_2024a}, we adopt an up-to-date, semi-empirical determination of the $Z$-dependent SFRD following \cite{Boco_2021}, with updates from more recent works. The main ingredients of this approach are the galaxy stellar mass functions (GSMFs), $\Phi(M_{\star})=d^2 N/dV\,d\log M_{\star}$, indicating the number of galaxies with a given stellar mass, $M_{\star}$, per unit comoving volume; the galaxy main sequence (MS), $\psi(M_{\star})$, a well-established empirical relation between stellar mass and SFR $\psi$ of star-forming galaxies; and the fundamental metallicity relation (FMR), $Z_{\rm FMR}(M_{\rm\star},\psi)$, that connects $M_{\star}$ and $\psi$ to the galaxy metallicity.

We adopt the GSMFs from \cite{Chruslinska_2019}, selecting the case with fixed low-mass end slope $\alpha_{\rm SMF}=-1.45$, and the galaxy MS from \cite{Popesso_2022}. Moreover, we consider a double-Gaussian distribution for $\psi$ as in \cite{Sargent_2012}, accounting for both MS and starbursts (SB), i.e. galaxies experiencing brief and intense episodes of star formation:
\begin{equation}
\begin{split}
    \frac{dp}{d\log \psi}(\psi,M_{\rm \star},z) &= \frac{f_{\rm MS}}{\sigma_{\rm MS}\sqrt{2\pi}}\:\exp\left[-\frac{\left(\log \psi-\langle \log \psi\rangle_{\rm MS}\right)^2}{2\sigma_{\rm MS}^2}\right]\\
    & + \frac{f_{\rm SB}}{\sigma_{\rm SB}\sqrt{2\pi}}\:\exp\left[-\frac{\left(\log \psi-\langle \log \psi\rangle_{\rm SB}\right)^2}{2\sigma_{\rm SB}^2}\right].
\end{split}
\end{equation}
Here, $\braket{\log \psi}_{\rm MS}$ indicates the MS, and $\sigma_{\rm MS}=0.188$ is the dispersion of the $\psi$ distribution around the MS. For SBs, instead, we have $\braket{\log\psi}_{\rm SB}=\braket{\log\psi}_{\rm MS}+0.59$, and $\sigma_{\rm SB}=0.243$. $f_{\rm MS}$ and $f_{\rm SB}$ are the relative fractions of MS and SB galaxies, with $f_{\rm MS}+f_{\rm SB}=1$. We adopt the $f_{\rm SB}$ from \cite{Chruslinska_2021}, as a function of $M_{\star}$ and $z$. 

Regarding the FMR, we employ the determination by \cite{Curti_2020}:
\begin{equation}
Z_{\rm FMR}(M_{\rm \star},\psi)=8.779-(0.31/2.1)\times\log\left(1+(M_{\rm \star}/M_0(\psi))^{-2.1}\right),
\end{equation}
with $M_0(\psi)=10^{10.11}\times\psi^{0.56}$. Moreover, we prescribe a log-normal distribution around the FMR, as:
\begin{equation}
\frac{dp}{d\log Z}(Z,Z_{\rm FMR}(M_{\rm \star},\psi))\propto \exp\left[-\frac{\left(\log Z-\log Z_{\rm FMR}(M_{\rm \star},\psi)\right)^2}{2\sigma_{\rm Z}^2}\right].
\end{equation}
$\sigma_{\rm Z}$ describes the dispersion of the galaxy metallicity distribution around the FMR. In this work, we fix it to $\sigma_{\rm Z}=0.15$ (but see Sec.~\ref{sec:pisn_detection} where we also consider $\sigma_{\rm Z}=0.35$).

By convolving the above quantities, we finally obtain our $Z$-dependent SFRD:
\begin{equation}
\begin{split}
\frac{d^3M_{\star}}{dtdVd\log Z}(Z,z) = & \int d\log M_{\rm \star}\frac{d^2N}{dVd\log M_{\rm \star}}(M_{\rm \star},z)\\
& \times \int d\log \psi\:\psi \frac{dp}{d\log \psi}(\psi,M_{\rm \star},z)\\
& \times \frac{dp}{d\log Z}(Z,Z_{\rm FMR}(M_{\rm \star},\psi))\\
\end{split}
\end{equation}

\end{document}